  \documentclass[journal,doublecolumn]{IEEEtran}

  \usepackage{mathtools,amsfonts,mathrsfs,amssymb,amsthm}
  \usepackage{graphicx}
  \DeclareGraphicsExtensions{.eps}
  \usepackage{float}
  \usepackage{color}
  \usepackage{cite}
  \usepackage{acronym}
  \usepackage{siunitx}
  \usepackage{subfig}
  \usepackage{wasysym}
  \usepackage[normalem]{ulem}  
  
  \restylefloat{figure} 

  \def\Xbo{\mathbf{X}}
  \def\Xcal{\mathcal{X}}

  \def\Rset{\mathbb{R}} 
  \def\Bbo{\mathbf{B}} 
  \def\Ncal{\mathcal{N}}
  \def\nth{\ensuremath{n^{th}}}
  \def\kth{\ensuremath{k^{th}}}
  \def\s{\mathrm{s}}
  \def\b{\mathrm{b}}
  \def\d{\mathrm{d}}
  \def\c{\mathrm{c}}
  \def\e{\mathrm{e}}
  \def\p{\mathrm{p}}
  \def\ind#1{1_{#1}}
  \def\onebo{\mathbf{1}}
  \def\zerobo{\mathbf{0}}
  
  \newcommand{\ds}[1]{\displaystyle{#1}} 
  
  \newtheorem{proposition}{Proposition}

  \newtheorem*{theorem*}{Theorem}
  \newtheorem{theorem}{Theorem}
  \newtheorem{definition}{Definition}
  \newtheorem{example}{Example}

  \acrodef{pda}[PDA]{Probabilistic Data Association}
  \acrodef{jpda}[JPDA]{Joint Probabilistic Data Association}
  \acrodef{mht}[MHT]{Multiple Hypothesis Tracking}
  \acrodef{rfs}[RFS]{Random Finite Set}
  \acrodef{fisst}[FISST]{Finite Set Statistics}
  \acrodef{pgfl}[p.g.fl.]{Probability Generating Functional}
  \acrodef{phd}[PHD]{Probability Hypothesis Density}
  \acrodef{cphd}[CPHD]{Cardinalized Probability Hypothesis Density}
  \acrodef{gm}[GM]{Gaussian Mixture}
  \acrodef{iid}[i.i.d.]{independent and identically distributed}
  \acrodef{wrt}[w.r.t.]{with respect to}
  \acrodef{resp}[resp.]{respectively}
  \acrodef{rsos}[RSOs]{Resident Space Objects}
  \acrodef{ospa}[OSPA]{Optimal Sub-Pattern Assignment}
  \acrodef{first}[FIRST]{Faculty-in-Residence Summer Term}
  \acrodef{map}[MAP]{Maximum A Posteriori}
  \acrodef{mc}[MC]{Monte Carlo}
  \acrodef{zip}[ZiP]{Zero-inflated Poisson}
  \acrodef{fov}[FoV]{Field of View}
  
  \title{Spawning Models for the CPHD Filter}

  \author
  {
    Daniel~S.~Bryant, Emmanuel~D.~Delande, Steven~Gehly, J\'{e}r\'{e}mie~Houssineau, Daniel~E.~Clark, Brandon~A.~Jones
    \thanks
    {
      D. S. Bryant, S. Gehly, and B. A. Jones are with the Department of Aerospace Engineering Sciences, University of Colorado Boulder, Boulder, CO 80309 USA (e-mail: daniel.bryant@colorado.edu; steven.gehly@colorado.edu; brandon.jones@colorado.edu)
    }
    \thanks
    {
      E. D. Delande, J. Houssineau, and D. E. Clark are with the School of Engineering and Physical Sciences, Heriot-Watt University (HWU), Edinburgh EH14 4AS, U.K. (e-mail: E.D.Delande@hw.ac.uk; j.houssineau@hw.ac.uk; d.e.clark@hw.ac.uk)
    }
  }
  
\begin{document}
  \maketitle

  \begin{abstract}
    In its classical form, the \ac{cphd} filter does not model the appearance of new targets through spawning, yet there are applications for which spawning models more appropriately account for newborn objects when compared to spontaneous birth models. In this paper, we propose a principled derivation of the \ac{cphd} filter with spawning from the \acl{fisst} framework. A \acl{gm} implementation of the \ac{cphd} filter with spawning is then presented, illustrated with three applicable spawning models on a simulated scenario involving two parent targets spawning a total of five objects. Results show that filter implementations with spawn models provide more accurate results when compared to a birth model implementation.
  \end{abstract}
  

  \IEEEpeerreviewmaketitle
  
  \acresetall 

  \section{Introduction}
    \IEEEPARstart{T}{he} goal of the multi-object estimation problem is to jointly estimate -- usually in the presence of clutter, data association uncertainty, and missed detections -- the time-varying number and individual states of targets evolving in a surveillance scene. Commonly known detection and tracking algorithms for the multi-object problem include  \ac{jpda} \cite{Fortmann_TE_1983} and \ac{mht} \cite{Reid_D_1979}. Relatively new is the multi-object filtering framework known as \ac{fisst} \cite{Mahler_RPS_2007_3, Mahler_RPS_2014}, based on a representation of the target population as a \ac{rfs}, a specific case of the more general concept of point process.

    Within the \ac{fisst} framework, the multi-target Bayes filter proposes an optimal solution to the multi-object estimation problem; it is, however, impractical in realistic applications due to its combinatorial complexity \cite{Mahler_RPS_2007_3}. Several approximations of the multi-target Bayes filter have been proposed to circumvent this intractability, including the \ac{phd} \cite{Mahler_RPS_2003} and the \ac{cphd}  \cite{Mahler_RPS_2007} filters. The \ac{phd} filter propagates the first-order factorial moment density, or \emph{intensity}, of the multi-target \ac{rfs}, representing the whole population of targets within the surveillance scene \cite{Mahler_RPS_2003}. While inexpensive, the \ac{phd} filter exhibits a high variability in the estimated target number \cite{Mahler_RPS_2014}. The \ac{cphd} filter \cite{Mahler_RPS_2007} addresses this issue by estimating the cardinality distribution of the multi-target \ac{rfs} in addition to its intensity. Unlike for the \ac{phd} filter, the initial presentation of the \ac{cphd} filter does not include a model for target spawning. Target spawning refers to instances where a parent target generates one or more daughter targets and where the daughter(s) usually remain(s) in close proximity to the parent for some amount of time following their appearance, e.g., a fighter jet that launches a missile.

    Though the \ac{cphd} filter's model for birth targets has the potential to address spawning targets \cite{Mahler_RPS_2014}, there may be cases where specific spawning models are more applicable. In the context of the tracking of \ac{rsos}, natural and artificial Earth orbiting satellites consisting of active spacecraft, decommissioned payloads, and debris, consider for example the deployment of CubeSats from a launch vehicle \cite{Swartwout_M_2011, Swartwout_M_2012} or fragmentation events caused by the unintentional \cite{Anselmo_L_2009} or intentional \cite{Johnson_NL_2008} collision of objects. Without spawning, the best option may be the use of diffuse birth regions, however, the volume of space to be filled requires a potentially intractable number of birth regions \cite{Jones_BA_2015_1}. To improve the \ac{cphd} filter's performance for space-object tracking, \cite{Jones_BA_2014} presented a measurement-based birth model that leverages an astrodynamics approach to track initialization for \ac{rsos}.  While such an approach may be effective for tracking spawned \ac{rsos}, a multi-target filter that correctly models the birth process for a given target is expected to provide better accuracy and faster confirmation of new objects.  The models proposed in this paper allow for the development of CPHD implementations used for RSO tracking applications with spawning.
    
    The incorporation of spawning models in the context of \ac{cphd} filtering has previously been explored in \cite{Lundgren_M_2013_1}, relying on an intuitive construction of the filtering equations related to the spawning models considered (Bernoulli or Poisson process) through a non-standard derivation procedure. In this paper, we propose expressions for the \ac{cphd} filter enhanced with various target spawning models through a standard derivation procedure within the \ac{fisst} framework specific to the considered spawning model (Bernoulli, Poisson, or zero-inflated Poisson process). To the best of our understanding, the derivation of the spawning terms in \cite{Lundgren_M_2013_1} relies on additional approximations and the approach does not lead to the same results as those presented here.
        
    The structure of this paper is as follows. Section~\ref{sec:background} presents the relevant background on point processes and functional differentiation, followed by key definitions and properties pertinent to our results. Section~\ref{sec:derivation} provides a detailed construction of the \ac{cphd} filter with target spawning, considering several models of spawning processes. Section~\ref{sec:simulation} demonstrates the proposed concepts through simulation example, and closing remarks are given in Section~\ref{sec:conclusion}. The proofs of the results in Section~\ref{sec:derivation} are given in the Appendix.

  \section{Background} \label{sec:background}
    In this section, we introduce the necessary background on point processes (Section~\ref{subsec:point_process}), on \ac{pgfl}s (Section~\ref{subsec:pgfl}), on functional differentiation (Section~\ref{subsec:differentiation}), and on a few properties from the application of differentiation in the context of point processes (Section~\ref{subsec:property}).

    \subsection{Point processes} \label{subsec:point_process}
      A point process on some space $\Xbo$ is a random variable whose number of elements \emph{and} element states, belonging to $\Xbo$, are random. In the context of multi-target tracking the population of targets is represented by a point process $\Phi$, on a single-target state space $\Xbo \subseteq \Rset^d$, whose elements describe individual target states. A realization of $\Phi$ is a vector of points $\varphi = (x_1,\ldots, x_{N})$ depicting a specific multi-target configuration, where $x_i \in \Xbo$ describes the $d$-component state of an individual target (position, velocity, etc.).
      
      A point process $\Phi$ is characterized by its probability distribution $P_{\Phi}$ on the measurable space $(\Xcal, \Bbo_{\Xcal})$, where $\Xcal = \bigcup_{n \geq 0} \Xbo^n$ is the point process state space, i.e., the space of all the finite vectors of points in $\Xbo$, and $\Bbo_{\Xcal}$ is the Borel $\sigma$-algebra on $\Xcal$ \cite{Stoyan_D_1995}. The probability distribution of a point process is defined as a symmetric function, so that the order of points in a realization is irrelevant for statistical purposes -- for example, realizations $(x_1, x_2)$ and $(x_2, x_1)$ are equally probable. In addition, if the probability distribution is such that the realizations are vectors of points that are pairwise distinct \emph{almost surely}, then the point process is called \emph{simple}. For the rest of the paper, all the point processes are assumed simple\footnote{An alternative construction of simple point processes as random objects whose realizations are \emph{sets} of points $\varphi = \{x_1,\ldots, x_{N}\}$, in which the elements are per construction \emph{unordered}, is also available in the literature \cite{Mahler_RPS_2007, Vo_BN_2005}. In this context, a point process is called a \ac{rfs}.}.
      
      The probability distribution $P_{\Phi}$ is characterized by its projection measures $P_{\Phi}^{(n)}$, for any $n \geq 0$. The \nth-order projection measure $P_{\Phi}^{(n)}$, for any $n \geq 1$, is defined on the Borel $\sigma$-algebra of $\Xbo^n$ and gives the probability for the point process to be composed of $n$ points, and the probability distribution of these points. By extension, $P_{\Phi}^{(0)}$ is the probability for the point process to be empty. For any $n \geq 0$, $J^{(n)}_{\Phi}$ denotes the \nth-order Janossy measure \cite[p. 124]{VereJones_D_2003}, and is defined as
      \begin{subequations}
	\begin{align}
	  J_{\Phi}^{(n)}(B_1 \times \ldots \times B_n) &= \sum_{\sigma(n)} P_{\Phi}^{(n)}(B_{\sigma_1} \times \ldots \times B_{\sigma_n})
	  \\
	  &= n! P_{\Phi}^{(n)}(B_1 \times \ldots \times B_n),
	\end{align}
      \end{subequations}
      where $B_i$ is in $\Bbo_{\Xbo}$, the Borel $\sigma$-algebra of $\Xbo$, $1 \leq i \leq n$, and where $\sigma(n)$ denotes the set of all permutations $(\sigma_1, \ldots, \sigma_n)$ of $(1, \ldots, n)$.
      
      The probability density $p_{\Phi}$ (\ac{resp} the \nth-order projection density $p_{\Phi}^{(n)}$, the \nth-order Janossy density $j_{\Phi}^{(n)}$) is the Radon-Nikodym derivative of the probability distribution $P_{\Phi}$ (\ac{resp} the \nth-order projection measure $P_{\Phi}^{(n)}$, the \nth-order Janossy measure $J_{\Phi}^{(n)}$) \ac{wrt} some reference measure. All these quantities provide equivalent ways to describe the point process $\Phi$. However, a measure-theoretical formulation provides a more general framework that is required to construct certain statistical properties on point processes that can be exploited for practical applications; a recent example is given in \cite{Delande_E_2014_4} for the construction of the regional statistics. For the sake of generality, the rest of the paper thus uses a measure-based description.
      
      Assuming that $f$ is a non-negative measurable function on $\Xcal$, then the integral of $f$ \ac{wrt} to the measure $P_\Phi$ can be written in the following ways:
      \begin{subequations} \label{eq:point_process_description}
	\begin{align}
	  P_{\Phi}(f) &= \int_{\Xcal} f(\varphi)P_{\Phi}(\d\varphi)
	  \\
	  &= \int_{\Xcal} f(\varphi)p_{\Phi}(\varphi)\d\varphi
	  \\
	  &= \sum_{n \geq 0} \int_{\Xbo^n} f(x_1,\ldots,x_n)P_{\Phi}^{(n)}(\d(x_1, \ldots, x_n))
	  \\
	  &= \sum_{n \geq 0} \mathrlap{\int_{\Xbo^n}}~~~ f(x_1,\ldots,x_n)p_{\Phi}^{(n)}(x_1, \ldots, x_n) \d x_1 \ldots \d x_n
	  \\
	  &= \sum_{n \geq 0} \frac{1}{n!} \int_{\Xbo^n} f(x_1,\ldots,x_n)J_{\Phi}^{(n)}(\d(x_1, \ldots, x_n))
	  \\
	  &= \sum_{n \geq 0} \frac{1}{n!} \mathrlap{\int_{\Xbo^n}}~~~ f(x_1,\ldots,x_n)j_{\Phi}^{(n)}(x_1, \ldots, x_n) \d x_1 \ldots \d x_n.	  
	\end{align}
      \end{subequations}
      Throughout this article the exploitation of the Janossy measures will be preferred, for they are convenient tools in the context of functional differentiation (see Section~\ref{subsec:differentiation}). For the sake of simplicity, domains of integration will be omitted when they refer to the full target state space $\Xbo$.
      
      The Janossy measures can also be used directly to exploit meaningful information on the point process $\Phi$. For example, central to this article is the extraction of the \emph{cardinality distribution} $\rho_{\Phi}$ of the point process, that describes the number of elements in the realizations of $\Phi$ (see Section~\ref{sec:derivation}):
      
      \begin{example}[Cardinality distribution]
	Consider the function $f_n$ defined as
	\begin{equation}
	  f_n(\varphi) =
	  \begin{dcases}
	    1, &|\varphi| = n,
	    \\
	    0, &\textrm{otherwise},
	  \end{dcases}
	\end{equation}
	where $|\varphi|$ denotes the size of the vector $\varphi$. The integral of $f_n$ \ac{wrt} to $P_{\Phi}$ yields the probability $\rho_{\Phi}(n)$ that a realization $\varphi$ of the point process $\Phi$ has size $n$ and we have, using Eq.~\eqref{eq:point_process_description} (see \cite[p.28]{Srinivasan_SK_2003}):
	\begin{subequations} \label{eq:point_process_cardinality}
	  \begin{align}
	    \rho_{\Phi}(n) &= P_{\Phi}(f_n)
	    \\
	    &= \int_{\Xbo^n} P_{\Phi}^{(n)}(\d(x_1, \ldots, x_n))
	    \\
	    &= \frac{1}{n!} \int_{\Xbo^n} J_{\Phi}^{(n)}(\d(x_1, \ldots, x_n)). 
	  \end{align}
	\end{subequations}
	The function $\rho_{\Phi}$ is called the cardinality distribution of the point process $\Phi$. Note that the \nth-order projection measure $P_{\Phi}^{(n)}$ (\ac{resp} the \nth-order Janossy measure $J_{\Phi}^{(n)}$) is not a probability measure, in the general case, for its integral over $\Xbo^n$ yields $\rho_{\Phi}(n)$ (\ac{resp} $n!\rho_{\Phi}(n)$).
      \end{example}  
          
    \subsection{Probability generating functionals} \label{subsec:pgfl}
      The \ac{pgfl} provides a useful characterization for point process theory \cite{Moyal_JE_1962} and is defined as follows.
      \begin{definition}[Probability generating functional \cite{VereJones_D_2003}] \label{def:pgfl}
	The probability generating functional $G_{\Phi}$  of a point process $\Phi$ on $\Xbo$ can be written for any test function $h \in \mathcal{U}(\Xbo)$ as\footnote{$\mathcal{U}(\Xbo)$ is the space of bounded measurable functions $u$ on $\Xbo$ satisfying $||u||_{\infty} \leq 1$.}
	\begin{subequations} \label{eq:point_process_pgfl}
	  \begin{align}
	    G_{\Phi}(h) &= \int_{\Xcal} \Big[\prod_{x \in \varphi} h(x)\Big] P_{\Phi}(\d\varphi) \label{eq:point_process_pgfl_1}
	    \\
	    &= J^{(0)}_{\Phi} ~\mathclap{+}~ \sum_{n\geq 1} \dfrac{1}{n!} \mathrlap{\int_{\Xbo^n}}~~~ h(x_1) \ldots h(x_n) J^{(n)}_{\Phi}(\d(x_1, \ldots, x_n)). \label{eq:point_process_pgfl_2}
	  \end{align}
	\end{subequations}
      \end{definition}
      The \ac{pgfl} $G_{\Phi}$ fully characterizes the point process $\Phi$, and is a very convenient tool for the extraction of statistical information on $\Phi$ through functional differentiation (see Section~\ref{subsec:differentiation}). From Eq.~\eqref{eq:point_process_pgfl} we can immediately write
      \begin{align}
	G_{\Phi}(0) &= J^{(0)}_{\Phi}~(= P^{(0)}_{\Phi}),
	\\
	G_{\Phi}(1) &= 1.
      \end{align}
      Operations on point processes (e.g., superposition of two populations) can be translated into operations on their corresponding \ac{pgfl}s. In the context of multi-target tracking, \ac{pgfl}s provide a convenient description of the compound population (targets or measurements) resulting from an operation on elementary populations.
      
      The superposition operation for point processes describes the union of two populations $\Phi_1$, $\Phi_2$ into a compound population $\Phi_1 \cup \Phi_2$, during which the information about the origin population of each individual is lost.
      \begin{proposition}[Superposition of independent processes \cite{Moyal_JE_1962}] \label{def:superposition}
	Let $\Phi_1$ and $\Phi_2$ be two independent point processes defined on the same space, with respective \ac{pgfl}s $G_{\Phi_1}$ and $G_{\Phi_2}$. The \ac{pgfl} of the superposition process $\Phi_1 \cup \Phi_2$ is given by the product
	\begin{equation}
	  G_{\Phi_1 \cup \Phi_2}(h) = G_{\Phi_1}(h)G_{\Phi_2}(h). \label{eq:superposition}
	\end{equation}
      \end{proposition}

      The Galton-Watson recursion for point processes \cite{Watson_HW_1875, Moyal_JE_1962} describes the evolution of each individual $x$ from a parent population $\Phi_{\p}$ into a population of daughter individuals, independently of the other parent individuals but following a common evolution model described by a process $\Phi_{\e}$. The resulting daughter population $\Phi_{\d}$ is then the superposition of all the populations of daughter individuals.
      \begin{proposition}[The Galton-Watson recursion \cite{Watson_HW_1875}] \label{def:galton_watson}
	Let $G_{\Phi_{\p}}$ be the \ac{pgfl} of a parent process $\Phi_{\p}$ on $\Xbo$, and let $G_{\Phi_{\e}}(\cdot|x)$ be the conditional \ac{pgfl} of an evolution process $\Phi_{\e}$, defined for every $x \in \Xbo$. The \ac{pgfl} of the daughter process $\Phi_{\d}$ is given by the composition
	\begin{equation}
	  G_{\Phi_{\d}}(h)= G_{\Phi_{\p}}\left(G_{\Phi_{\e}}(h|\cdot)\right). \label{eq:galton_watson}
	\end{equation}
      \end{proposition}

    \subsection{Functional differentiation} \label{subsec:differentiation}
      To make use of functionals in the derivations presented in Section~\ref{sec:derivation}, we require the notion of differentials on functional spaces. We adopt a restricted form of the G\^ateaux differential, known as the chain differential \cite{Bernhard_P_2006}, so that a general chain rule can be determined \cite{Clark_DE_2013_2, Clark_DE_2013_3}. Following this, we describe the general higher-order chain rule.

      \begin{definition}[Chain differential \cite{Bernhard_P_2006}]
	Under the conditions detailed in \cite{Bernhard_P_2006}, the function $f$ on some set $X$ has a {\it chain differential} $\delta f(x;\eta)$ at $x \in X$ in the direction $\eta$ if, for any sequence $\eta_n\rightarrow\eta\in X$, and any sequence of real numbers $\theta_n\rightarrow 0$, it holds that
	\begin{equation}
	  \delta f(x;\eta) = \lim_{n\rightarrow \infty} \dfrac{1}{\theta_n} \left( f(x+\theta_n\eta_n)-f(x) \right). \label{eq:functional_derivative}
	\end{equation}
      \end{definition}

      The \nth-order chain differential  can be defined recursively as
      \begin{equation}
	\delta^n f\left(x;\eta_1,\ldots,\eta_n\right) = \delta\left(\delta^{n-1} f\left(x;\eta_1,\ldots,\eta_{n-1}\right); \eta_n\right).
      \end{equation}

      Applying \nth-order chain differentials on composite functions can be an extremely laborious process since it involves determining the result for each choice of function and proving the result by induction. For ordinary derivatives, the general higher-order chain rule is normally attributed to Fa\`a di Bruno \cite{FaaDiBruno_F_1855}. The following result generalizes Fa\`a di Bruno's formula to chain differentials and allows for a systematic derivation of composite functions (see \cite{Clark_DE_2013_2} for an example of exploitation in the context of Bayesian estimation).
      
      \begin{proposition}[General higher-order chain rule, from \cite{Clark_DE_2013_3, Clark_DE_2015_1}] \label{prop:chain_rule}
	Under the differentiability and continuity conditions detailed in \cite{Clark_DE_2015_1}, the \nth-order variation of composition $f\circ g$ in the sequence of directions $(\eta_i)_{i=1}^n$ at point $x$ is given by
	\begin{multline}
	  \delta^n (f \circ g)(x; (\eta_i)_{i=1}^n)
	  \\
	  = \sum_{\pi\in \Pi_n} \delta^{|\pi|}f \bigg(g(x); \Big(\delta^{|\omega|} g \big(x; (\eta_i)_{i \in \omega}\big)\Big)_{\omega \in \pi} \bigg), \label{eq:chain_rule}
	\end{multline}
	where $\Pi_n = \Pi(\{1,\ldots,n\})$ represents the set of partitions of the index set $\{1,\ldots,n\}$, and $|\pi|$ denotes the cardinality of the set $\pi$.
      \end{proposition}
      
      \begin{example}[General higher-order chain rule]
	\begin{align}
	  &\delta^2 (f \circ g)(x; \eta_1,\eta_2) \nonumber
	  \\
	  &= \underbrace{\delta^2f\left(g(x); \delta g(x; \eta_1), \delta g(x; \eta_2)\right)}_{\pi = \{\{1\}, \{2\}\}} + \underbrace{\delta f\left(g(x);\delta^2 g(x; \eta_1, \eta_2)\right)}_{\pi = \{\{1, 2\}\}}.
	\end{align}
      \end{example}
      
      Applying \nth-order chain differentials on a product of functions follows a more straightforward approach, similar to Leibniz' rule for ordinary derivatives.
      
      \begin{proposition}[Leibniz' rule, from \cite{Clark_DE_2015_1}] \label{prop:product_rule}
	Under the differentiability conditions detailed in \cite{Clark_DE_2015_1}, the \nth-order variation of the product $f \cdot g$ in the sequence of directions $(\eta_i)_{i=1}^n$ at point $x$ is given by
	\begin{multline}
	  \delta^n (f \cdot g)(x; (\eta_i)_{i=1}^n)  
	  \\
	  = \sum_{\pi \subseteq  \{1, \ldots, n\}} \delta^{|\pi|} f(x; \left(\eta_i\right)_{i \in \pi}) \delta^{n-|\pi|} g(x; \left(\eta_i\right)_{i \in \pi^c}), \label{eq:product_rule}
	\end{multline}
	where $\pi^c = \{1, \ldots, n\} \setminus \pi$ denotes the complement of $\pi$ in $\{1, \ldots, n\}$.
      \end{proposition}
      
      \begin{example}[Leibniz' rule]
	\begin{multline}
	  \delta^2 (f \cdot g)(x; \eta_1,\eta_2) =
	  \\
	  \underbrace{\delta^2f(x;\eta_1,\eta_2)g(x)}_{\pi = \{1, 2\}} + \underbrace{\delta f(x;\eta_1)\delta g(x;\eta_2)}_{\pi = \{1\}}
	  \\
	  + \underbrace{\delta f(x;\eta_2)\delta g(x;\eta_1)}_{\pi = \{2\}} + \underbrace{f(x)\delta g(x;\eta_1,\eta_2)}_{\pi = \{\emptyset\}}.
	\end{multline}
      \end{example}
    \subsection{Probability generating functionals and differentiation} \label{subsec:property}
      Key properties of a point process can be recovered from the functional differentiation of its \ac{pgfl}. Taking the \kth-order variation of $G_{\Phi}(h)$ in the directions $\eta_1, \ldots, \eta_k$, we have (see, for example \cite[p. 21]{Srinivasan_SK_1973}),
      \begin{align}
	&\delta^k G_{\Phi}(h;{\eta_1,\ldots,\eta_k}) = \nonumber
	\\
	&\sum_{n \geq k} \dfrac{1}{(n-k)!} \int_{\Xbo^n} ~~\mathclap{\prod_{i=1}^k}~~ \eta_i(x_i) ~~~\mathclap{\prod_{i=k+1}^n}~~ h(x_i)~ J^{(n)}_\Phi(\d(x_1, \ldots, x_n)).
      \end{align}
      It is then useful to consider the cases when we set $h = 1$ or $h = 0$, i.e.
      \begin{align} 
	\delta^k G_\Phi(0&; \eta_1,\ldots,\eta_k) \nonumber
	\\
	&= \int_{\Xbo^k} \eta_1(x_1)\ldots\eta_k(x_k)J_\Phi^{(k)}(\d(x_1, \ldots, x_k)), \label{eq:pgfl_derivation_generic_0}
	\\
	\delta^k G_\Phi(1&; \eta_1,\ldots,\eta_k) \nonumber
	\\
	&= \int_{\Xbo^k} \eta_1(x_1)\ldots\eta_k(x_k)M_\Phi^{(k)}(\d(x_1, \ldots, x_k)), \label{eq:pgfl_derivation_generic_1}
      \end{align}
      where $M_\Phi^{(k)}$ is the \kth-order factorial moment measure, defined as in \cite[p. 111]{Stoyan_D_1995}.
%
      
      Assuming that one wishes to evaluate the Janossy and factorial moment measures in some measurable subsets $B_i \in \Bbo_{\Xbo}$, $1\leq i \leq k$, then they can be recovered from Eqs~\eqref{eq:pgfl_derivation_generic_0}, \eqref{eq:pgfl_derivation_generic_1} by setting the directions to be indicator functions\footnote{For a measurable subset $B \in \Bbo_{\Xbo}$, the indicator function $\ind{B}$ is defined as the function on $\Xbo$ such that $\ind{B}(x) = 1$ if $x \in B$, $\ind{B}(x) = 0$ otherwise.} $\eta_i = \ind{B_i}$, $1\leq i \leq k$, so that
      \begin{align}
	\left.\delta^k G_\Phi(h;\ind{B_1},\ldots,\ind{B_k})\right|_{h = 0} &= J_\Phi^{(k)}(B_1 \times \ldots \times B_k), \label{eq:pgfl_derivation_setting_0}
	\\
	\left.\delta^k G_\Phi(h;\ind{B_1},\ldots,\ind{B_k})\right|_{h = 1} &= M_\Phi^{(k)}(B_1 \times \ldots \times B_k). \label{eq:pgfl_derivation_setting_1}
      \end{align}
      
      The propagation of the first-order factorial moment measure $M_\Phi^{(1)}$ -- also called the \emph{intensity measure} $\mu_{\Phi}$ -- of the multi-target point process $\Phi$, in a Bayesian context, is a key component of the construction of both the \ac{phd} filter \cite{Mahler_RPS_2003} and the \ac{cphd} filter \cite{Mahler_RPS_2007}. The density of the intensity measure is called the \emph{Probability Hypothesis Density} \cite{Mahler_RPS_2003}.

  \section{The CPHD filter with spawning} \label{sec:derivation}
    This section covers the derivation of the filtering equations for the \ac{cphd} filter for various target spawning processes. Section~\ref{subsec:cphd_general} provides a brief description of the general multi-target Bayes filter \cite{Mahler_RPS_2007_3}, and the principled approximation leading to the construction of the original \ac{cphd} filter \cite{Mahler_RPS_2007}. Section~\ref{subsec:process_model} then presents the various models of point processes that will be necessary for the construction of the \ac{cphd} filter with spawning in Section~\ref{subsec:cphd_prediction_step}.
    
    \subsection{Multi-object filtering and CPHD filter} \label{subsec:cphd_general}
      The multi-target Bayes filter \cite{Mahler_RPS_2007_3} is the natural extension of the usual single-target Bayesian paradigm to the multi-target case, within the \ac{fisst} framework. The multi-target Bayes recursion at time step $k$ consists of the \emph{time prediction} and \emph{data update} steps given as follows:
      \begin{align}
	\!\!P_{k|k-1}(\d\varphi|Z_{1:k-1}) &= \!\int_{\Xcal} \!\!\!f_{k|k-1}(\varphi|\bar{\varphi})P_{k-1}(\d\bar{\varphi}|Z_{1:k-1}),
	\\
	\!\!P_k(\d\varphi|Z_{1:k}) &= \!\frac{g_k(Z_k|\varphi)P_{k|k-1}(\d\varphi|Z_{1:k-1})}{\int_{\Xcal} g_k(Z_k|\bar{\varphi})P_{k|k-1}(\d\bar{\varphi}|Z_{1:k-1})},
      \end{align}
      where $P_{k|k-1}$ (\ac{resp} $P_k$) is the probability distribution of the predicted multi-target process $\Phi_{k|k-1}$ (\ac{resp} the posterior multi-target process $\Phi_k$), $Z_i$, $1 \leq i \leq k$, is the set of measurements collected at time step $i$, $Z_{1:i}$ denotes the sequence $Z_1,\ldots,Z_i$, $f_{k|k-1}$ is the multi-target transition kernel, and $g_k$ is the multi-target likelihood function. The multi-target transition kernel $f_{k|k-1}$ describes the time evolution of the population of targets since time step $k-1$ and encapsulates the underlying models of target birth, motion, spawning, and death. The multi-target likelihood $g_k$ describes the sensor observation process and encapsulates the underlying models of target detection, target-generated measurements, and false alarms.
      
      The multi-target Bayes recursion is used to propagate the posterior distribution $P_k(\cdot|Z_{1:k})$ that describes the current target population based on all the measurements $Z_1,\ldots,Z_k$ collected so far. The \ac{cphd} Bayes recursion aims at simplifying the multi-target Bayes recursion by approximating the predicted and posterior multi-target processes as \ac{iid} processes\footnote{The definition of an \ac{iid} process is given in Section~\ref{subsec:process_model}.}, a class of point processes fully characterized by their cardinality distribution $\rho_{\Phi}$ and their first-order moment measure $\mu_{\Phi}$ \cite{Mahler_RPS_2007}. The \ac{cphd} filter thus focuses on the propagation of the posterior cardinality distribution $\rho_k$ and the posterior first-order moment measure $\mu_k$, rather than the full posterior probability distribution $P_k$.
      
      
      The original construction of the \ac{cphd} filter \cite{Mahler_RPS_2007} does not consider a target spawning mechanism, and the key contribution of this paper is to propose the integration of several target spawning models in the \ac{cphd} time prediction equation (see Section~\ref{subsec:cphd_prediction_step}). Note that the data update step does not involve the target spawning mechanism and is therefore left out of the scope of this paper. A detailed description of the data update step can be found in \cite{Vo_BT_2007}.
    
    \subsection{Point process models} \label{subsec:process_model}
      \subsubsection{Bernoulli process}
	A Bernoulli process $\Phi$ is characterized by a parameter $0 \leq p \leq 1$ and a spatial distribution $s$. It describes the situation where 1) either there is no object in the scene, or 2) there is a single object in the scene, with state distributed according to $s$. Its projection measures are given by
	\begin{equation}
	  P_{\Phi}^{(n)}(B_1 \times \ldots \times B_n) =
	  \begin{cases}
	    1 - p, &n = 0,
	    \\
	    ps(B_1), &n = 1,
	    \\
	    0, &\text{otherwise.}
	  \end{cases}
	\end{equation}
    
	\begin{proposition} [\ac{pgfl} of a Bernoulli process \cite{Mahler_RPS_2014}] \label{lem:pgfl_bernoulli}
	  The \ac{pgfl} of a Bernoulli process $\Phi$ with parameter $p$ and spatial distribution $s$ is given by
	  \begin{equation}
	    G_{\Phi}(h) = 1 - p + p \int h(x) s(\d x). \label{eq:pgfl_bernoulli}
	  \end{equation}  
	\end{proposition}
	
      \subsubsection{Poisson process}
	A Poisson process $\Phi$ is characterized by a rate $\lambda \geq 0$ and a spatial distribution $s$. It describes a population whose size follows a Poisson distribution and whose individual states are \ac{iid} according to $s$. Its projection measures are given by
	\begin{equation}
	  P_{\Phi}^{(n)}(B_1 \times \ldots \times B_n) = e^{-\lambda}\frac{\lambda^n}{n!} \ds{\prod_{i = 1}^n} s(B_i).
	\end{equation}
	
	\begin{proposition} [\ac{pgfl} of a Poisson process \cite{Mahler_RPS_2014}] \label{lem:pgfl_poisson}
	  The \ac{pgfl} of a Poisson process $\Phi$ with rate $\lambda$ and spatial distribution $s$ is given by
	  \begin{equation} 
	    G_{\Phi}(h) = \exp\left[\lambda\left(\int h(x)s(\d x) - 1\right)\right]. \label{eq:pgfl_poisson}
	  \end{equation}  
	\end{proposition}

      \subsubsection{Zero-inflated Poisson process}
	A zero-inflated Poisson process $\Phi$ (from \cite{Lambert_D_1992}) is characterized by a parameter $0 \leq p \leq 1$, a rate $\lambda \geq 0$, and a spatial distribution $s$. It describes a population that is 1) either empty, or 2) non-empty, with size following a Poisson distribution and whose individual states are \ac{iid} according to $s$. Its projection measures are given by
	\begin{equation}
	  P_{\Phi}^{(n)}(B_1 \times \ldots \times B_n) =
	  \begin{cases}
	    1 - p + pe^{-\lambda}, &n = 0,
	    \\
	    pe^{-\lambda}\frac{\lambda^n}{n!} \ds{\prod_{i = 1}^n} s(B_i), &\text{otherwise.}
	  \end{cases}
	\end{equation}
	Note that a Poisson process is a special case of a zero-inflated Poisson process in which the parameter $p$ is set to one.
	
	\begin{proposition} [\ac{pgfl} of a zero-inflated Poisson process]\label{lem:pgfl_bernoulli_poisson}
	  The \ac{pgfl} of a zero-inflated Poisson process $\Phi$ with parameter $p$, rate $\lambda$, and spatial distribution $s$ is given by
	  \begin{equation}
	    G_\Phi(h) = 1 - p + p\exp\left[\lambda\left(\int h(x)s(\d x) - 1\right)\right]. \label{eq:pgfl_bernoulli_poisson}
	  \end{equation}  
	\end{proposition}

      \subsubsection{I.i.d. process}
	An \ac{iid} process $\Phi$ is characterized by a cardinality distribution $\rho$ and a spatial distribution $s$. It describes a population whose size is distributed according to $\rho$, and whose individual states are \ac{iid} according to $s$. Its Janossy measures are given by
	\begin{equation}
	  J_{\Phi}^{(n)}(B_1 \times \ldots \times B_n) = n!\rho(n)\ds{\prod_{i = 1}^n} s(B_i). \label{eq:janossy_iid}
	\end{equation}
	Note that a Poisson process is a special case of \ac{iid} process in which the cardinality distribution $\rho$ is Poisson.
	
    \subsection{Prediction step} \label{subsec:cphd_prediction_step}
      In this section, we propose an alternative expression of the original \ac{cphd} time prediction step \cite{Mahler_RPS_2007} in which newborn targets originate from a spawning mechanism rather than spontaneous birth. Note that the assumptions on the posterior multi-target process from the previous time step, the target survival mechanism, and the target evolution mechanism are identical to the original assumptions in \cite{Mahler_RPS_2007}.
    
      \begin{theorem}[CPHD with spawning: prediction step] \label{th:prediction_step}
	Assuming that, at step $k$:
	\begin{itemize}
	  \item The posterior multi-target process $\Phi_{k-1}$ is an \ac{iid} process with intensity measure $\mu_{k-1}$, with cardinality distribution $\rho_{k-1}$, and spatial distribution $s_{k-1}$, 
	  \item A target in state $x$ at time $k - 1$ survived to time $k$ with probability $p_{\s, k}(x)$,
	  \item A surviving target in state $x$ at time $k - 1$ evolved since time $k - 1$ according to a Markov transition $f_{\s, k}(\cdot|x)$,
	  \item There was no spontaneous target birth since time $k - 1$,
	  \item Newborn targets were spawned from prior targets (see next page),
	\end{itemize}
	then the intensity measure $\mu_{k|k-1}$ and cardinality distribution $\rho_{k|k-1}$ of the predicted multi-target process $\Phi_{k|k-1}$ are given by
	\begin{align}
	  \mu_{k|k-1}(\cdot) &= \int \left[p_{\s, k}(x)f_{\s, k}(\cdot|x) + \mu_{\b, k}(\cdot|x)\right]\mu_{k-1}(\d x), \label{eq:prediction_intensity}
	  \\
	  \rho_{k|k-1}(n) &= \sum_{j=1}^n B_{n,j}(b_1,\ldots,b_n) \nonumber
	  \\
	  &\quad\quad\times \left[\sum_{m \geq j} \frac{m!}{n!(m-j)!}\rho_{k-1}(m){b_0}^{m-j}\right], \label{eq:prediction_cardinality}
	\end{align}
	where $B_{n,j}$ is the partial Bell polynomial \cite{Charalambides_C_2002} given by
	\begin{multline}
	  B_{n,k}(x_1,x_2,\cdots,x_n) =
	  \\
	  \mathrlap{\sum_{\substack{\vspace{3pt} \\ k_1 + 2k_2 + \cdots + nk_n = n \\ k_1 + k_2 + \cdots + k_n = k}}}\hspace{50pt} \frac{n!}{k_1!(1!)^{k_1} k_2!(2!)^{k_2} \cdots k_n!(n!)^{k_n}} x_1^{k_1} x_2^{k_2} \cdots x_n^{k_n}, \label{eq:BellPoly}
	\end{multline}
	and where the intensity measure $\mu_{\b, k}$ and the coefficients $b_i$ are the parameters of the spawning process, dependent on the modeling choices. Denoting $\bar{p}_{\s, k}(\cdot) \equiv 1 - p_{\s, k}(\cdot)$ and $\bar{p}_{\b, k}(\cdot) \equiv 1 - p_{\b, k}(\cdot)$, the parameters are as follows:

	a) Bernoulli process, with parameter $p_{\b, k}$ and spatial distribution $s_{\b, k}$:
	\begin{equation}
	  \mu_{\b, k}(\cdot|x) = p_{\b, k}(x)s_{\b, k}(\cdot|x),
	\end{equation}
	and
	\begin{equation}
	  b_i =
	  \begin{cases}
	    \int \bar{p}_{\s, k}(x)\bar{p}_{\b, k}(x)s_{k-1}(\d x), &i = 0,
	    \\
	    \int \left[p_{\s, k}(x)\bar{p}_{\b, k}(x) + \bar{p}_{\s, k}(x)p_{\b, k}(x)\right]s_{k-1}(\d x), &i = 1,
	    \\
	    2\int p_{\s, k}(x)p_{\b, k}(x) s_{k-1}(\d x), &i = 2,
	    \\
	    0, &i > 2.
	  \end{cases}
	\end{equation}
	
	b) Poisson process, with rate $\lambda_{\b, k}$ and spatial distribution $s_{\b, k}$:
	\begin{equation}
	  \mu_{\b, k}(\cdot|x) = \lambda_{\b, k}(x)s_{\b, k}(\cdot|x),
	\end{equation}
	and,
	\begin{multline}
	  b_i = \int \lambda_{\b, k}^{i-1}(x) e^{-\lambda_{\b, k}(x)} 
	  \\
	  \times \left[\bar{p}_{\s, k}(x)\lambda_{\b, k}(x) + ip_{\s, k}(x)\right]s_{k-1}(\d x)
	  \\
	  i\geq 0,
	\end{multline}

	c) zero-inflated Poisson process, with parameter $p_{\b, k}$, rate $\lambda_{\b, k}$, and spatial distribution $s_{\b, k}$:
	\begin{equation}
	  \mu_{\b, k}(\cdot|x) = p_{\b, k}(x)\lambda_{\b, k}(x)s_{\b, k}(\cdot|x),
	\end{equation}
	and,
	\begin{equation}
	  b_i =
	  \begin{cases}
	    \int \bar{p}_{\s, k}(x)\left[\bar{p}_{\b, k}(x) + p_{\b, k}(x)e^{-\lambda_{\b, k}(x)}\right]s_{k-1}(\d x),
	    \\
	    \hspace{200pt}i = 0,
	    \\
	    \int \big[\bar{p}_{\s, k}
	    \\
	    \hspace{20pt}+ p_{\s, k}(x)\left[\bar{p}_{\b, k}(x) + p_{\b, k}(x)e^{-\lambda_{\b, k}(x)}\right]\big]s_{k-1}(\d x),
	    \\
	    \hspace{200pt}i = 1,
	    \\
	    \int p_{\b, k}(x)\lambda_{\b, k}^{i-1}(x) e^{-\lambda_{\b, k}(x)}
	    \\
	    \hspace{60pt}\times\left[\bar{p}_{\s, k}(x)\lambda_{\b, k}(x) + i p_{\s, k}(x)\right]s_{k-1}(\d x),
	    \\
	    \hspace{200pt}i \geq 2.
	  \end{cases}
	\end{equation}
      \end{theorem}
      The proof is given in the Appendix. Note that the structure of the predicted cardinality \eqref{eq:prediction_cardinality} allows for its efficient computation through an algorithm dedicated to the computation of partial Bell polynomials (see \cite{Charalambides_C_2002},\cite{Cvijovic_D_2011_1} for examples).

  \section{Simulation} \label{sec:simulation}
    In this section we illustrate the \ac{cphd} filter with spawning models through a simulation-based scenario. The \ac{gm} implementation of the \ac{cphd} filter is briefly described in Section~\ref{subsec:GaussMixImpl}, followed by a description of the metrics exploited for the analysis of the filter results in Section~\ref{subsec:metric}. The scenario and the selection of the filter parameters are detailed in Section~\ref{subsec:Scenario}, and the results are discussed in Section~\ref{subsec:Results}.

    \subsection{The \ac{gm}-\ac{cphd} filter with spawning} \label{subsec:GaussMixImpl}
      Since the incorporation of spawning in the \ac{cphd} filtering process does not affect the data update step, we shall focus in this section on the specifics of the prediction step for the \ac{gm}-\ac{cphd} filter with spawning. A description of the usual \ac{gm}-\ac{cphd}, including the implementation of the spontaneous birth term, is given in \cite{Vo_BT_2007}.
	
      \subsubsection{Filtering assumptions} \label{subsubsec:assumption}
	We follow the usual assumptions of the \ac{gm}-\ac{cphd} filter \cite{Vo_BT_2007} regarding the transition process from time $k-1$ to time $k$, namely, that the probability of survival $p_{\s, k}$ is uniform over the state space $\Xbo$ and the transition $f_{\s, k}$ follows a linear Gaussian dynamical model:
	\begin{align}
	  p_{\s, k}(\cdot) &\equiv p_{\s, k}, \label{eq:prediction_survival}
	  \\
	  f_{\s, k|k-1}(\cdot|x) &= \Ncal(\cdot~; F_k x, Q_k), \label{eq:prediction_motion}
	\end{align}
	where $\Ncal(\cdot~; m, P)$ denotes a Gaussian distribution with mean $m$ and covariance $P$, $F_k$ is a state transition matrix, and $Q_k$ is a process noise covariance matrix.
	
	Regardless of the chosen spawning model (see Theorem~\ref{th:prediction_step}), we further assume that the spatial distribution of each spawned object $s_{\b, k}$ can be described as the Gaussian mixture
	\begin{equation}
	  s_{\b,k}(\cdot|x) = \sum_{j=1}^{J_{\b,k}} w_{\b,k}^{(j)} \Ncal(\cdot~;F_{\b,k}^{(j)}x + d_{\b,k}^{(j)}, Q_{\b,k}^{(j)}), \label{eq:spawning_distribution}
	\end{equation}
	where $d_{\b,k}^{(j)}$ is a deviation vector, $F_{\b,k}^{(j)}$ is a spawning transition matrix, and $Q_{\b,k}^{(j)}$ is a spawning noise covariance matrix, for $1 \leq j \leq J_{\b,k}$, and $\sum_{j=1}^{J_{\b,k}} w_{\b,k}^{(j)} = 1$. Also, we assume that the model parameters $p_{\b, k}$, $\lambda_{\b, k}$, when applicable, are uniform over the state space $\Xbo$:
	\begin{equation}
	  \begin{aligned}
	    p_{\b, k}(\cdot) &\equiv p_{\b,k},
	    \\
	    \lambda_{\b, k}(\cdot) &\equiv \lambda_{\b,k}.
	  \end{aligned}
	\end{equation}
	    
      \subsubsection{Predicted intensity}
	The construction of the predicted intensity $\mu_{k|k-1}$ in Eq.~\eqref{eq:prediction_intensity} follows a similar structure as for the usual \ac{gm}-\ac{cphd} filter \cite{Vo_BN_2006_2}. Assume that the posterior intensity $\mu_{k-1}$ can be written as a Gaussian mixture of the form
	\begin{equation}
	  \mu_{k-1}(\cdot) = \sum_{j=1}^{J_{k-1}} w_{k-1}^{(j)} \Ncal(\cdot~; m_{k-1}^{(j)}, P_{k-1}^{(j)}),
	\end{equation}
	where $m_{k-1}^{(j)}$ (\ac{resp} $P_{k-1}^{(j)}$) is the posterior mean (\ac{resp} covariance) of the $j$-th component of the mixture. Then the predicted intensity $\mu_{k|k-1}$ can also be written as a Gaussian mixture of the form
	\begin{equation}
	  \mu_{k|k-1}(\cdot) = \mu_{\s,k|k-1}(\cdot) + \mu_{\b,k|k-1}(\cdot),
	\end{equation}
	where the surviving component $\mu_{\s, k|k-1}$ is the Gaussian mixture
	\begin{equation}
	  \mu_{\s,k|k-1}(\cdot) = p_{\s,k}\sum_{j=1}^{J_{k-1}} w_{k-1}^{(j)} \Ncal(\cdot~; m_{\s,k|k-1}^{(j)},P_{\s,k|k-1}^{(j)}),
	\end{equation}
	with
	\begin{align}
	  m_{\s,k|k-1}^{(j)} &= F_k m_{k-1}^{(j)},
	  \\
	  P_{\s,k|k-1}^{(j)} &= Q_k + F_k P_{k-1}^{(j)}F_k^T,
	\end{align}
	for $1 \leq j \leq J_{k-1}$, and the spawning component $\mu_{\b, k|k-1}$ is the Gaussian mixture
	\begin{multline} \label{eq:SpawnIntensPredict}
	  \mu_{\b, k|k-1}(\cdot)
	  \\
	  = \alpha_{\b,k} \sum_{j=1}^{J_{k-1}} w_{k-1}^{(j)} \sum_{i=1}^{J_{\b,k}} w_{\b,k}^{(i)} \Ncal(\cdot~; m_{b,k|k-1}^{(j, i)},P_{b,k|k-1}^{(j, i)}),
	\end{multline}
	with
	\begin{align}
	  m_{\b,k|k-1}^{(j, i)} &= F_{\b,k}^{(i)}m_{k-1}^{(j)} + d_{\b,k}^{(i)},
	  \\
	  P_{\b,k|k-1}^{(j, i)} &= Q_{\b,k}^{(i)} + F_{\b,k}^{(i)} P_{k-1}^{(j)} (F_{\b,k}^{(i)})^T,
	\end{align}
	for $1 \leq j \leq J_{k-1}$, $1 \leq i \leq J_{\b,k}$, and the scalar $\alpha_{\b,k}$ depends on the spawning model:
	\begin{equation}
	  \alpha_{\b,k} =
	  \begin{dcases}
	    p_{\b,k}, &\textrm{Bernoulli process},
	    \\
	    \lambda_{\b,k}, &\textrm{Poisson process},
	    \\
	    p_{\b,k}\lambda_{\b,k}, &\textrm{zero-inflated Poisson process}.
	  \end{dcases}
	\end{equation}
	
      \subsubsection{Predicted cardinality distribution} 
	Due to the assumptions presented in Section~\ref{subsubsec:assumption}, the coefficients of the Bell polynomial in Eq.~\eqref{eq:prediction_cardinality} have the simpler form
	\begin{align}
	  \intertext{a) Bernoulli process:}
	  b_i &=
	  \begin{cases} \label{eq:BernCardPredictIntmd}
	    \left(1 - p_{\s, k}\right)\left(1 - p_{\b, k}\right), &i = 0,
	    \\
	    p_{\s, k}\left(1 - p_{\b, k}\right) + \left(1 - p_{\s, k}\right)p_{\b, k}, &i = 1,
	    \\
	    2 p_{\s, k} p_{\b, k}, &i = 2,
	    \\
	    0, &i > 2.
	  \end{cases}
	  \intertext{b) Poisson process:}
	  b_i &= \lambda_{\b, k}^{i-1} e^{-\lambda_{\b, k}} \left[\left(1 - p_{\s, k}\right)\lambda_{\b, k} + ip_{\s, k}\right], \quad i \geq 0. \label{eq:PoissCardPredictIntmd}
	  \intertext{c) zero-inflated Poisson process:}
	  b_i &=
	  \begin{cases} \label{eq:BernPoissCardPredictIntmd}
	    \left(1 - p_{\s, k}\right)\left(1 - p_{\b, k} + p_{\b, k}e^{-\lambda_{\b, k}}\right), &i = 0,
	    \\
	    \left(1 - p_{\s, k}\right)p_{\b, k} e^{-\lambda_{\b, k}}\lambda_{\b, k} &
	    \\
	    \hspace{55pt}+ p_{\s, k}\left(1 - p_{\b, k} + p_{\b, k} e^{-\lambda_{\b, k}}\right), &i = 1,
	    \\
	    p_{\b, k}\lambda_{\b, k}^{i-1} e^{-\lambda_{\b, k}}\left[\left(1 - p_{\s, k}\right)\lambda_{\b, k} + i p_{\s, k}\right], &i \geq 2.
	  \end{cases}
	\end{align}
	The predicted cardinality distribution is then computed by the appropriate substitution of Eqs.~\eqref{eq:BernCardPredictIntmd}-\eqref{eq:BernPoissCardPredictIntmd} into Eq.~\eqref{eq:prediction_cardinality}.
	 	  
    \subsection{Evaluation metrics} \label{subsec:metric}
      To compare the multi-target state representing the true targets in the scene -- the ``ground truth'' -- and a collection of targets extracted from the filter's output, we exploit the \ac{ospa} metric \cite{Schuhmacher_D_2008} for assessing the accuracy of multi-object filters. Given two sets $X = \{x_1, \ldots, x_m\}$, $x_i \in \Xbo$, $1 \leq i \leq m$, and $Y = \{y_1, \ldots, y_n\}$, $y_j \in \Xbo$, $1 \leq j \leq n$, the second-order \ac{ospa} distance $d^{(c)}_2(X, Y)$ between $X$ and $Y$ is defined as
      \begin{multline}
	d^{(c)}_2(X, Y) =
	\\
	\begin{dcases}
	  0, &m = n = 0,
	  \\
	  \biggl[\!\frac{1}{n}\biggl(\!\min_{\pi \in \Pi_n}\!\sum_{i = 1}^m \!d^{(c)}(x_i, y_{\pi(i)})^2\!+\! c^2(n-m)\!\biggr)\!\biggr]^{1/2}\!\!\!\!\!\!\!,\!\!\!\!\! &m \leq n,
	  \\
	  d^{(c)}_2(Y, X), &\textrm{otherwise},
	\end{dcases}
      \end{multline}
      with
      \begin{equation}
	d^{(c)}(x_i, y_j) = \min(c, ||x_i - y_j||),
      \end{equation}
      where $c$ is the cutoff parameter, and $||\cdot||$ is the usual norm on $\Xbo$. The \ac{ospa} distance is such that $0 \leq d^{(c)}_2(X, Y) \leq c$; $d^{(c)}_2(X, Y) = 0$ indicates that $X$ and $Y$ are identical, while $d^{(c)}_2(X, Y)$ increases with the discrepancies between $X$ and $Y$, taking into account mismatches in number of elements and element states.
      
      In order to compare the true number of targets in the scene and a estimated cardinality distribution extracted from the filter's output, we exploit the Hellinger distance \cite{Gibbs_AL_2002_1}. Given two finite cardinality distributions $P= (p_1,\ldots, p_k)$ and $Q=(q_1,\ldots,q_k)$, the Hellinger distance $d_H(P,Q)$ is
      \begin{equation}
      \label{eq:Hell}
	d_H(P,Q) = \frac{1}{\sqrt{2}}\sqrt{\sum_{i = 1}^k (\sqrt{p_i} - \sqrt{q_i})^2}.
      \end{equation}
      Note that in \eqref{eq:Hell}, the coefficient $1/ \sqrt{2}$ is included in order to scale the Hellinger distance such that it is bounded as $0 \leq d_H(P,Q) \leq 1$; $d_H(P,Q) = 0$ indicates that $P$ and $Q$ are equivalent, where as $d_H(P,Q) \rightarrow 1$, $P$ and $Q$ become increasingly dissimilar.	
	      
    \subsection{Scenario and filter setup} \label{subsec:Scenario}
\begin{figure}[H]
	\centering
	\includegraphics[scale=0.5]{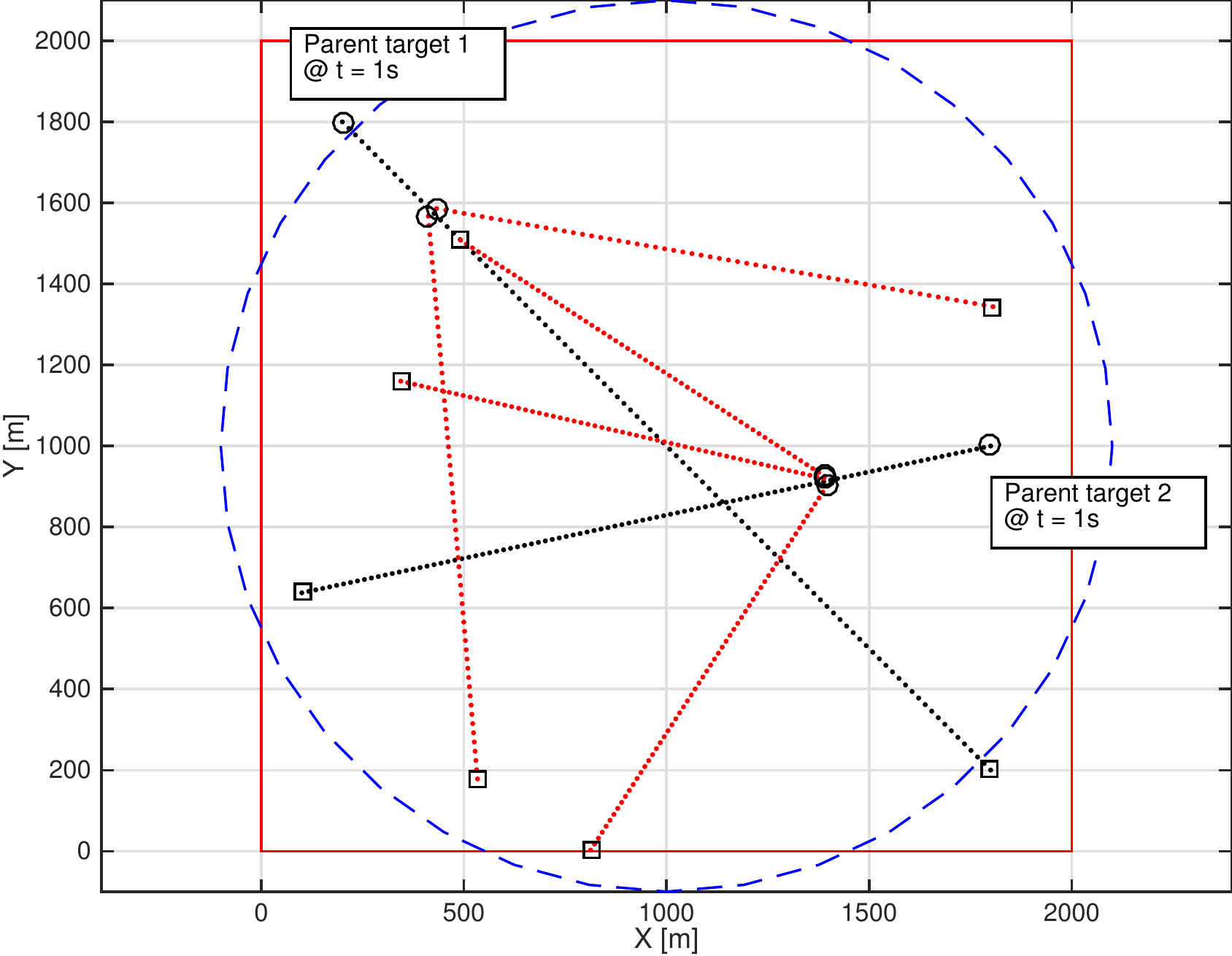} 
	\caption{Target trajectories. A circle ``\(\Circle \)'' indicates where a trajectory begins, and a square ``\( \square \)'' indicates where a trajectory ends. The large square indicates the limits of the sensor's \ac{fov} and the large dashed circle represents the $90\%$ confidence region of the Gaussian component of the spontaneous birth model.\label{fig:Scenario}}
      \end{figure}

\begin{figure}[H] 
	\centering
	\begin{tabular}{c}
	\subfloat[$x$-axis]{\includegraphics[scale=0.45]{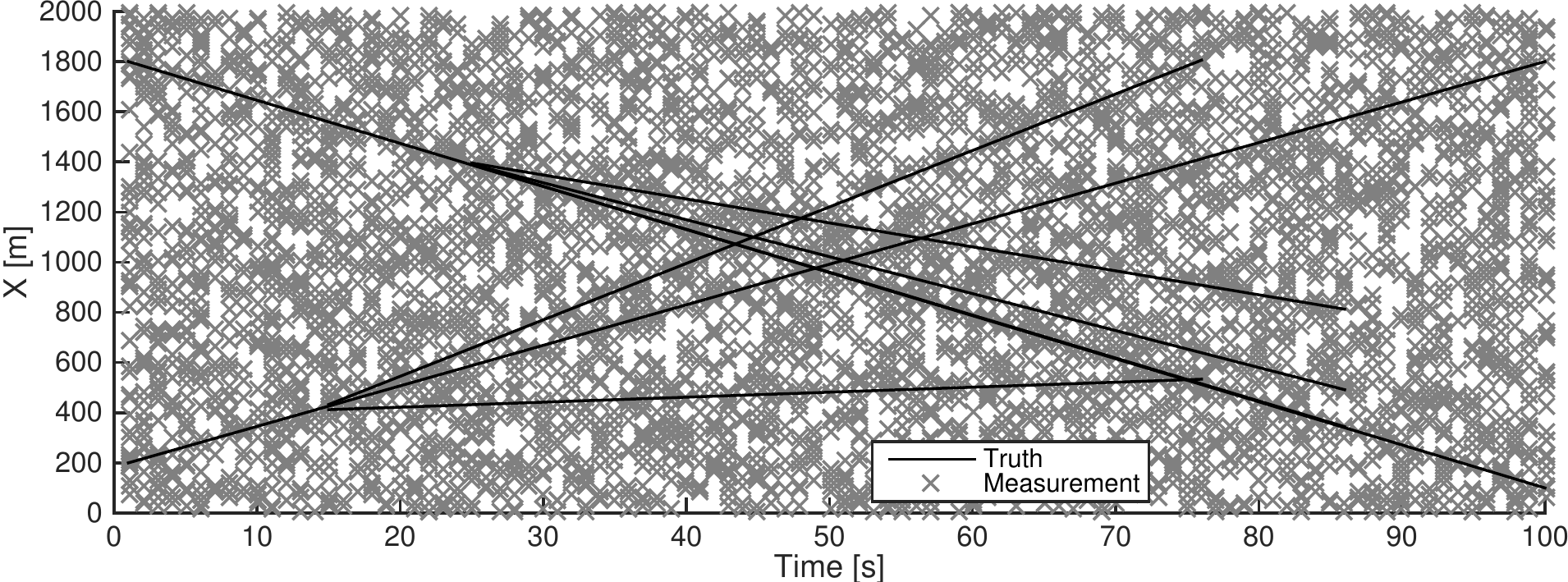}}
	\\
	\subfloat[$y$-axis]{\includegraphics[scale=0.45]{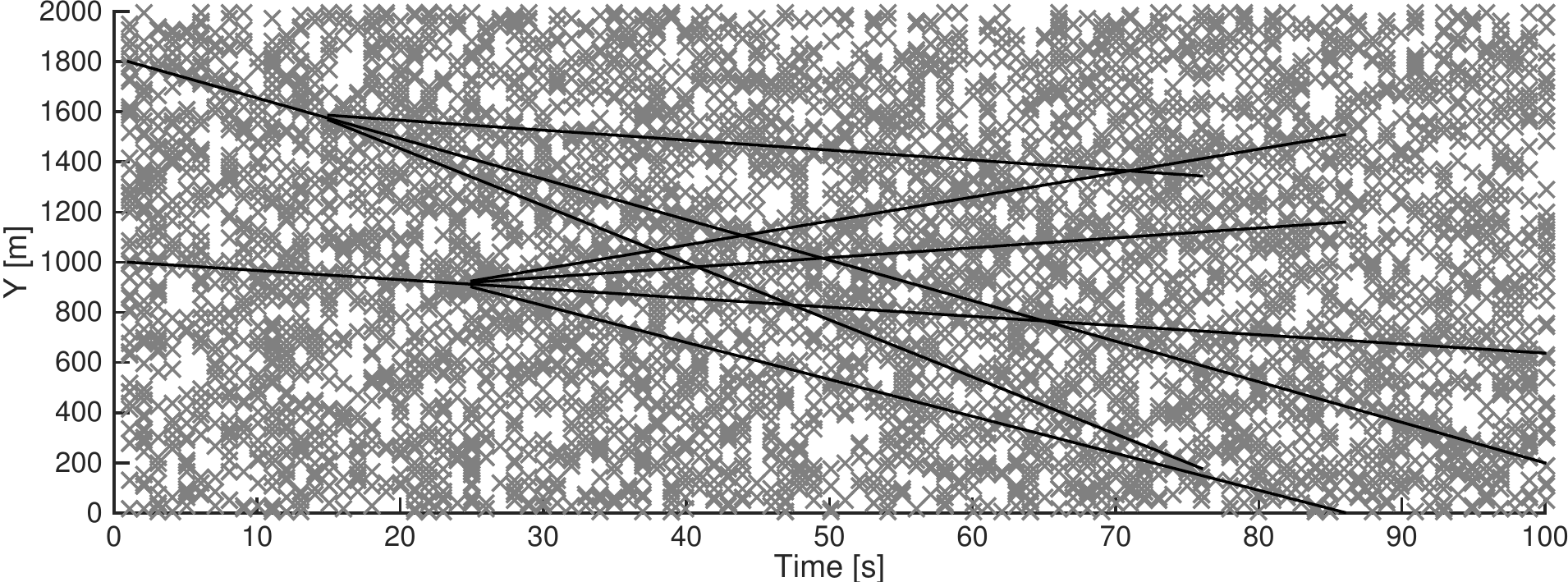}} 
	\end{tabular}
	\caption{Collected measurements (gray crosses) and target positions (black lines).\label{fig:Trajectories}}
      \end{figure}

      A point $[x, y, \dot{x}, \dot{y}]$ of the single-target state space $\Xbo \subset \Rset^4$ describes the position and velocity coordinates of an object in a square surveillance region of size $\SI{2000}{\meter}\times\SI{2000}{\meter}$. The simulated multi-target tracking scenario consists of  one scan per second for $\SI{100}{\second}$, and up to seven targets evolving in the region with constant velocity. Two targets are present at the beginning of the scenario and each spawns targets at different times: target $1$ spawns two additional targets at $t = \SI{15}{\second}$ and target $2$ spawns three additional targets at $t = \SI{25}{\second}$. All spawned targets have a lifespan of $\SI{60}{\second}$. Fig.~\ref{fig:Scenario} shows the trajectories of the targets cumulated over time, while Fig.~\ref{fig:Trajectories} illustrates these trajectories and the collected measurements across time.
	  
      The probability of survival $p_{\s,k}$ \eqref{eq:prediction_survival} is constant throughout the scenario, and set to $p_{\s,k} = 0.99$. The target motion model $f_{\s, k|k-1}$ \eqref{eq:prediction_motion} is set as follows:
      \begin{equation}
	F_k =
	\begin{bmatrix}
	  \onebo_2 & \Delta \onebo_2
	  \\
	  \zerobo_2 & \onebo_2 
	\end{bmatrix}
	,\quad\quad Q_k = \sigma_\nu^2
	\begin{bmatrix}
	  \frac{\Delta^4}{4} \onebo_2 & \frac{\Delta^3}{2} \onebo_2
	  \\
	  \frac{\Delta^3}{2} \onebo_2 & \Delta^2 \onebo_2 
	\end{bmatrix},
      \end{equation}
      where $\Delta = \SI{1}{\second}$, $\sigma_\nu = \SI{5}{\meter\second\tothe{-2}}$, and $\onebo_n$ (\ac{resp} $\zerobo_n$) denotes the $n\times n$ identity (\ac{resp} zero) matrix.
      
      The sensor's probability of detection is uniform over the sensor's \ac{fov}, and set at a constant value of $0.95$ throughout the scenario. Each target-generated measurement consists of the target's coordinate position with an independent Gaussian white noise on each component, with a standard deviation of $\SI{10}{\meter}$. Spurious measurements are modeled as a Poisson point process with uniform spatial distribution over the state space and an average number of clutter per unit volume of $\SI{12.5e-6}{\meter\tothe{-2}}$, that is, an average of $50$ clutter returns per scan over the surveillance region.
	  
      For the sake of comparison, the usual \ac{gm}-\ac{cphd} filter \cite{Vo_BT_2007} with spontaneous birth and no spawning is implemented as well. The spontaneous birth model is Poisson, with a constant rate of $0.025$ per time step (which yields, over the $\SI{100}{\second}$ of the scenario, an average of $2.5$ newborn targets for each parent target). The spatial distribution is modeled with a single Gaussian component, centered on the sensor's \ac{fov} as illustrated in Fig.~\ref{fig:Scenario}.
      

      The spatial distribution of the spawning \eqref{eq:spawning_distribution} is identical for the three considered models. We assume no spawned target deviation vectors, and a standard deviation of $12$ units is set on each component of the spawning noise covariance, i.e.
      \begin{equation}
	F_{\b, k} = 
	\begin{bmatrix}
	  \onebo_2 & \zerobo_2
	  \\
	  \zerobo_2 & \onebo_2 
	\end{bmatrix}
	,~ d_{\b, k} = \mathbf{0}
	,~ Q_{\b, k} =
	\begin{bmatrix}
	  \sigma_{\b}^2\onebo_2 & \zerobo_2
	  \\
	  \zerobo_2 & \dot{\sigma}_{\b}^2\onebo_2
	\end{bmatrix},
      \end{equation}
      where $\mathbf{0}$ denotes the null vector in $\Xbo$, $\sigma_{\b} = \SI{12}{\meter}$, and $\dot{\sigma}_{\b} = \SI{12}{\meter\second\tothe{-1}}$. 

      The parameters of the three spawning models are set as follows. The zero-inflated Poisson model assumes one spawning per parent target during the scenario with an average of $2.5$ daughter targets per spawning event, thus $p_{\b,k}$ and $\lambda_{\b}$ are set to $0.01$ and $2.5$, respectively. Relative to the zero-inflated Poisson model, the Poisson model is set to yield a similar spawning intensity thus its $\lambda_{\b,k}$ is set to $0.025$, whereas the Bernoulli model is set to yield a similar spawning frequency so its $p_{\b,k}$ is set to $0.01$. These parameters are also presented in Table \ref{tab:spawning}.
      
      \begin{table}[H]
	\caption{Spawn model parameters. \label{tab:spawning}}
	\centering
	\begin{tabular}{cccc}
	  \hline 
	  Model & $p_{\b, k}$ & $\lambda_{\b, k}$ & $\mu_{\b, k}(\cdot|x)$
	  \\
	  \hline
	  Bernoulli & $0.01$ & - & \;$0.01\Ncal(\cdot~; x, Q_{\b, k})$
	  \\
	  Poisson & - & $0.025$ & $0.025\Ncal(\cdot~; x, Q_{\b, k})$
	  \\
	  zero-inflated Poisson & $0.01$ & $2.5$ & $0.025\Ncal(\cdot~; x, Q_{\b, k})$
	\end{tabular}
      \end{table}
      
      It is interesting to note that neither the Poisson nor the Bernoulli models are equipped to capture the nature of the spawning events occurring in this scenario, since, per construction, the Poisson model is a poor match for spawning events occuring at unknown dates and the Bernoulli model is a poor match for spawning events creating more than one daughter target. The zero-inflated Poisson model possesses a greater flexibility and should be able to cope with a wider range of spawning situations; in any case, it is expected to yield better performances on the scenario presented in this paper.
 
      To maintain tractability, \ac{gm} components are truncated with threshold $T = 10^{-5}$, pruned with maximum number of components $J_{\max} = 100$, and merged with threshold $U = 4$ (see \cite{Vo_BN_2006_2} for more details on the pruning and merging mechanisms). Additionally, the maximum number of targets is set to $N_{\max} = 20$ to circumvent issues with infinitely tailed cardinality distributions \cite{Vo_BT_2007}.

    \subsection{Simulation results} \label{subsec:Results}      
      The proposed spawning models and the birth model are implemented with the \ac{gm}-\ac{cphd} filter, and compared over $500$ \ac{mc} runs of the multi-target scenario decribed in Section~\ref{subsec:Scenario}.
      
      \begin{figure}[H]
	\centering
	\includegraphics[scale=0.47]{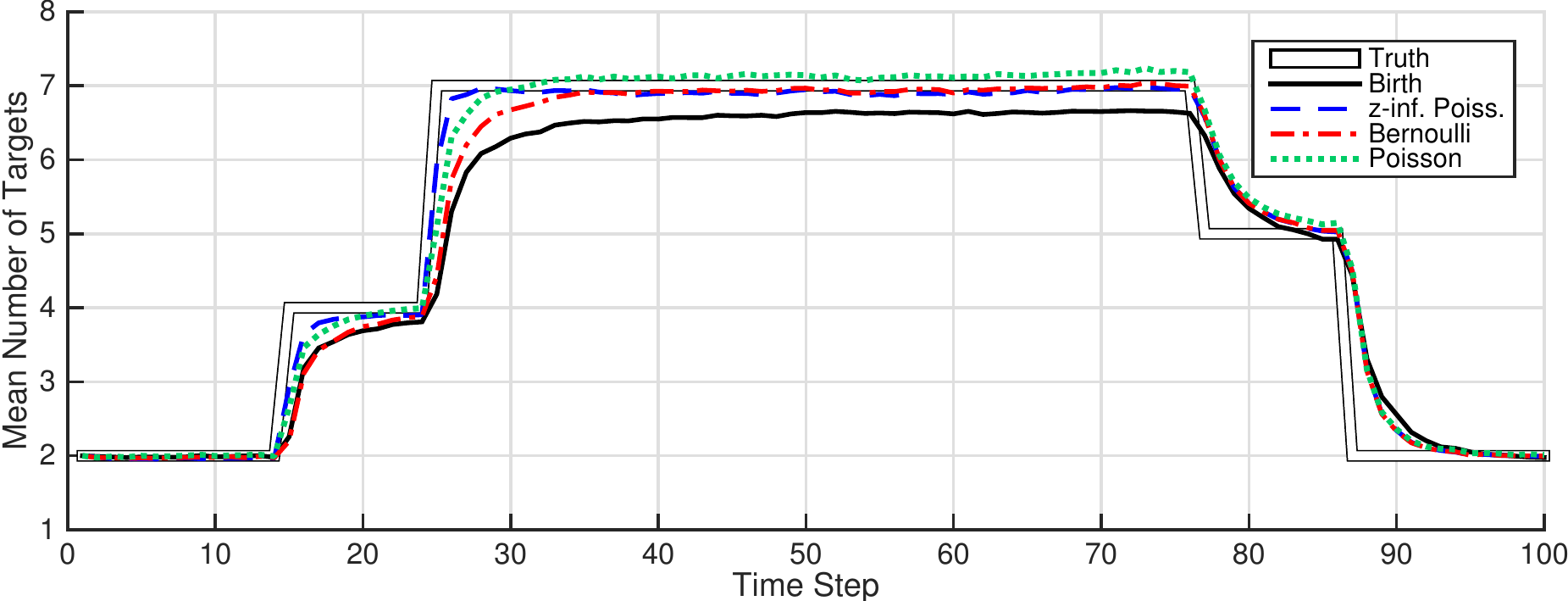} 
	\caption{\ac{map} estimate of the number of targets (averaged on $500$ runs).}
	\label{fig:CardEst}
      \end{figure}
      
      The \ac{map} estimate of the number of targets is plotted in Fig.~\ref{fig:CardEst}, along with the true number of targets in the scene. The results suggest that the spawning models provide a better estimate of the number of targets and, in particular, converge faster to the true number of targets following the appearance of new targets in the scene. This is expected, because the scenario does not feature any spontaneous but only spawning-related births, and thus in this context spawning models are a better match than the birth model.
      
      Among the three spawning models, the zero-inflated Poisson converges the fastest following the appearance of new targets, while the Bernoulli model converges the slowest. This is expected, for the zero-inflated Poisson model provides the best match to the spawning events occurring in this scenario. Note in particular that the Bernoulli model may not consider the appearance of more than one daughter per spawning event, and must therefore stage the multiple-target appearances across several successive time steps; in other words, the Bernoulli is ill-adapted to ``busy'' events where targets appear simultaneously. Note also the slight overestimation shown by the Poisson model when the true number of target is stable. Per construction, the Poisson model is well-equipped for the simultaneous appearance of an arbitrary number of spawned targets at any time step, but it fails at coping with ``quiet'' periods where no spawning occurs because, unlike the zero-inflated Poisson model, it does not temper the Poisson-driven spawning with a probability of spawning. In other words, the Poisson model is ill-adapted to the spawning events shown in this scenario.
      
      Note that all models -- spawning and birth -- follow the same mechanism for target deaths and yield much closer performances when target disappearances occur. 

      \begin{figure}[H] 
	\centering
	\begin{tabular}{c}
	  \subfloat[Position]{\includegraphics[scale=0.46]{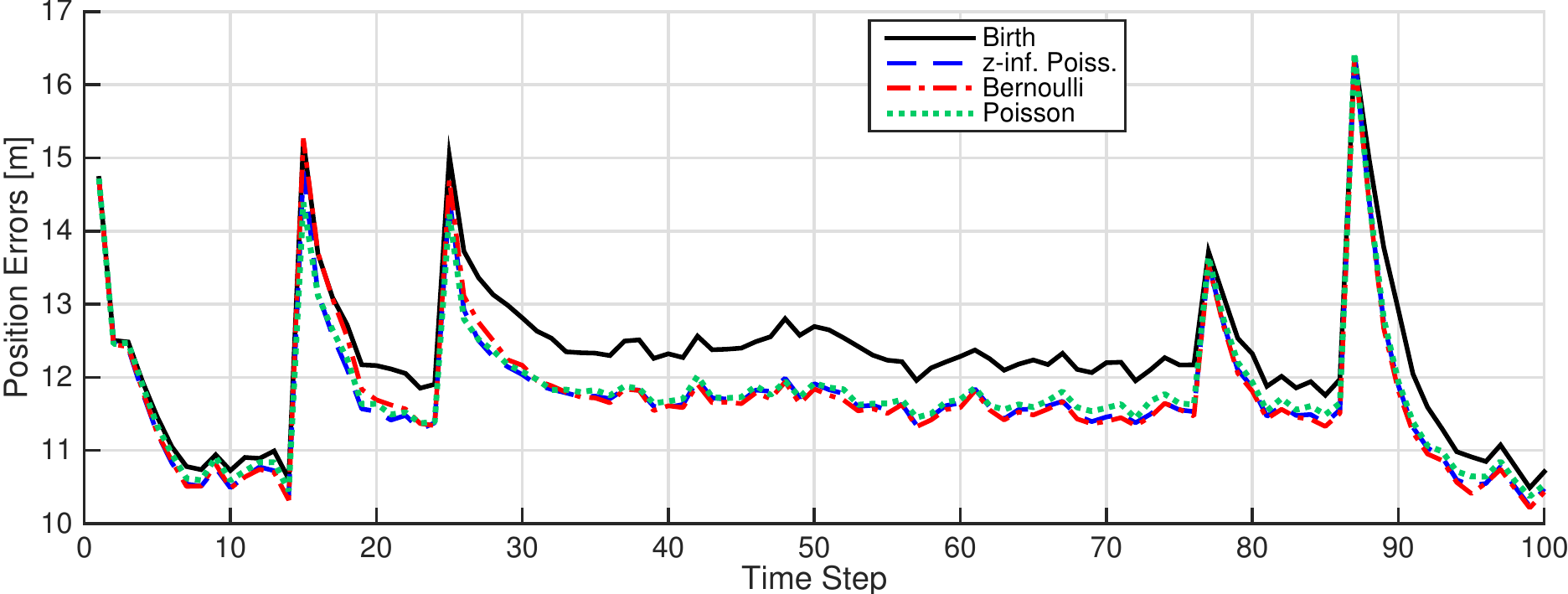}}
	  \\
	  \subfloat[Velocity]{\includegraphics[scale=0.46]{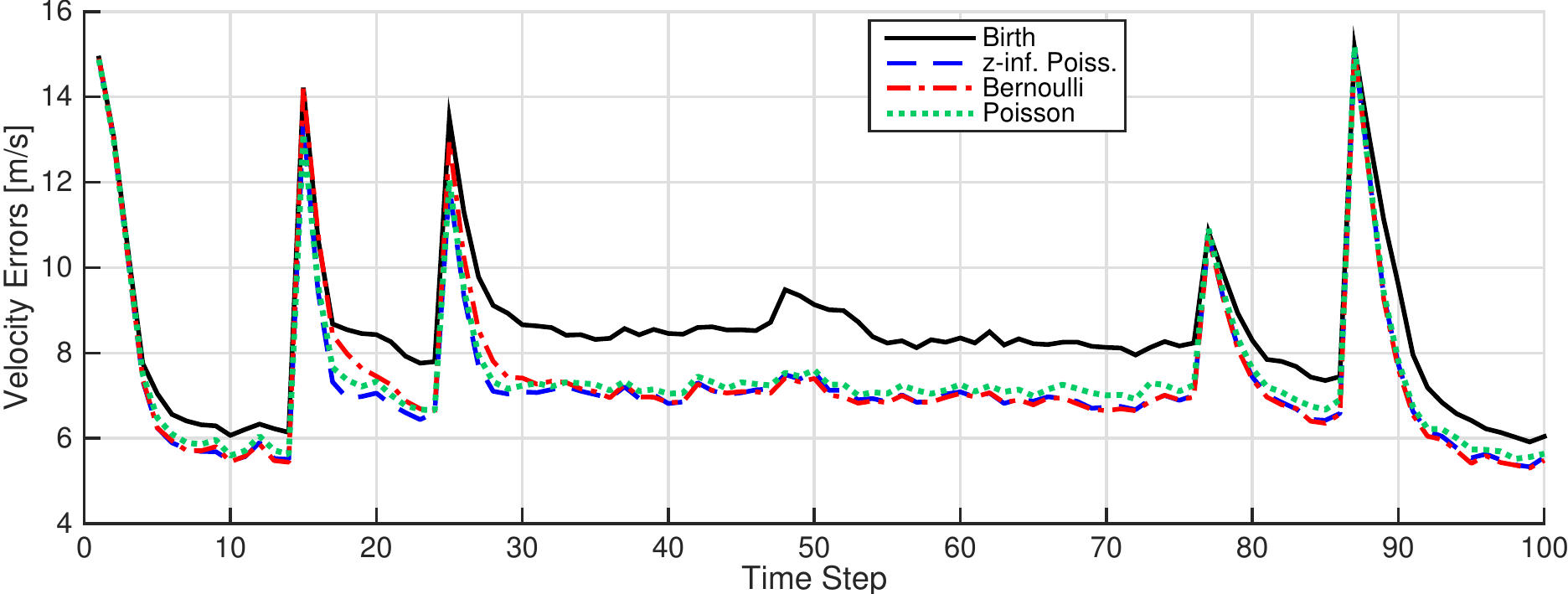}} 
	\end{tabular}
	\caption{\ac{ospa} distance (averaged on $500$ runs).}
	\label{fig:OSPA}
      \end{figure}
      
      Similar conclusions can be drawn from the comparisons of the \ac{ospa} distances shown in Fig.~\ref{fig:OSPA}. All models show error spikes at times of spawning ($t=\SI{15}{\second}$, $t=\SI{25}{\second}$) and death ($t=\SI{76}{\second}$, $t=\SI{86}{\second}$), however, the spawning models recover more quickly than the birth model, and have consistently lower errors.
      
      \begin{figure}[H] 
	\centering
	\begin{tabular}{c}
	  \subfloat[Predicted cardinality]{\includegraphics[scale=0.45]{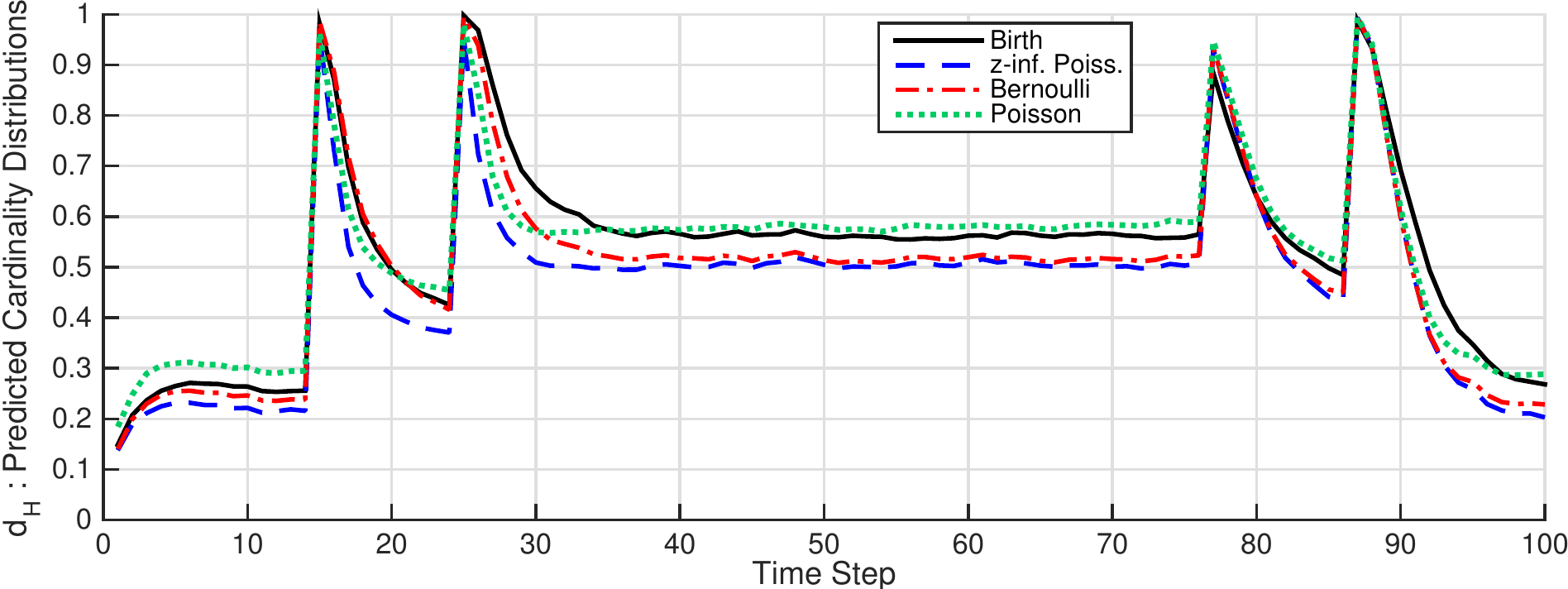}}
	  \\
	  \subfloat[Updated cardinality]{\includegraphics[scale=0.45]{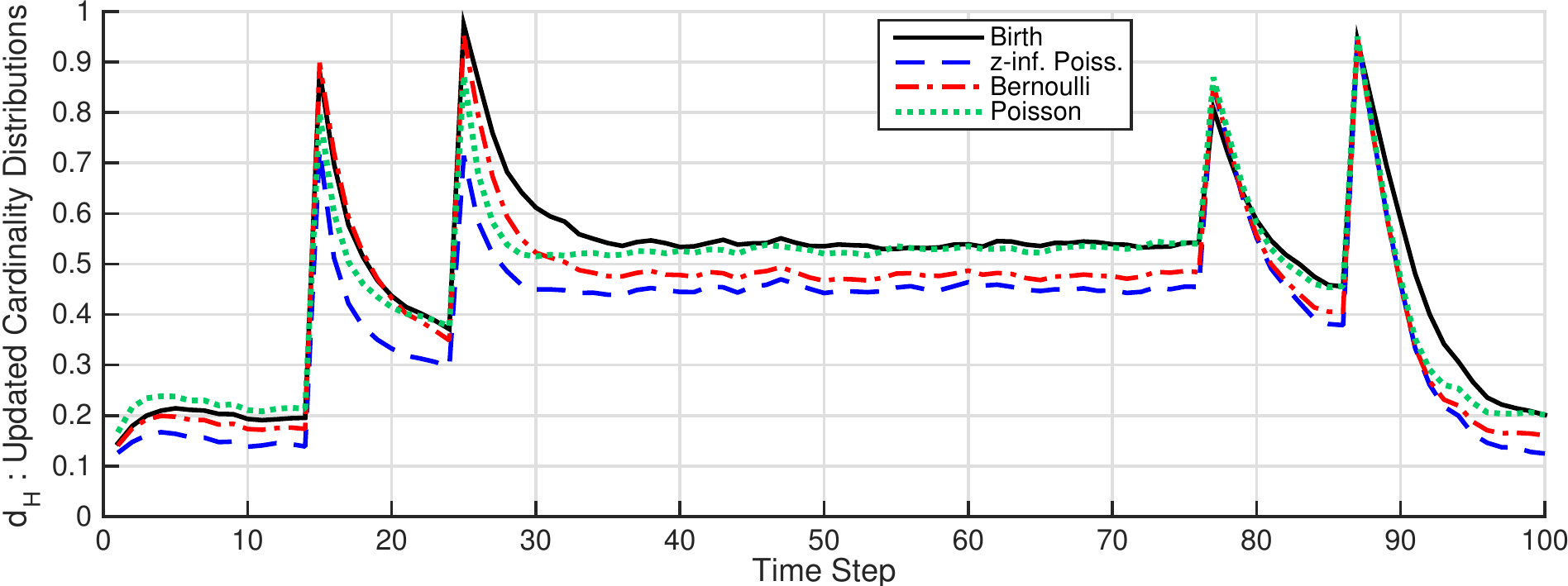}} 
	\end{tabular}
	\caption{Hellinger distances (averaged on $500$ runs).}
	\label{fig:Hellinger}
      \end{figure}
      
      The quality of the estimation of the number of targets proposed by the four models is further illustrated in Fig.~\ref{fig:Hellinger}, where the Hellinger distance between the cardinality distribution propagated by each model and the ``ideal'' cardinality distribution (i.e., a distribution in which all the mass is concentrated on the true number of targets).

      The results in Fig.~\ref{fig:Hellinger} allow a more refined analysis of the proposed models. All the models yield poor estimates immediately after a change in the true number of targets \footnote{Recall from Eq.~\eqref{eq:Hell} that the Hellinger distance $d_H$ is such that $0 \leq d_H \leq 1$.}, but the zero-inflated Poisson model converges the fastest following a target birth/death \emph{and} it converges to the best estimate during periods where the number of target is stable. The Poisson model converges \emph{faster} than the Bernoulli model, but to a \emph{worse} estimate: this is expected, since the Poisson model is ill-adapted to ``quiet'' periods while the Bernoulli model is ill-adapted to ``busy'' events (see discussion above on Fig.~\ref{fig:CardEst}).
      
      As expected, the updated cardinality distributions are consistently more accurate than the predicted cardinality distributions since they benefit from the processing of an additional measurement batch.
%
%
%

  \section{Conclusion} \label{sec:conclusion}
    The motivation for the work presented in this paper is the resolution of multi-object detection and tracking problems in which newborn objects are 
    spawned from preexisting ones. To this end, the construction of a \ac{cphd} filter in which the appearance of newborn targets is modeled with a spawning mechanism rather than spontaneous birth is proposed, based on a principled derivation procedure within the \ac{fisst} framework.

    A \ac{gm} implementation of the \ac{cphd} filter with spawning is then presented, considering three different models for the spawning mechanism based on a Bernoulli, a Poisson, or a zero-inflated Poisson process. The three resulting filters are then illustrated, analyzed, and compared to a usual \ac{cphd} filter with spontaneous birth but no spawning, on the same simulated scenario involving two parent targets spawning a total of five daughter targets. Results show that a spawning model, appropriately chosen for a given application, can provide better estimates than a spontaneous birth model.
    

  \section*{Acknowledgement}
    Daniel Bryant's work is supported by the Science, Mathematics \& Research for Transformation (SMART) Scholarship-for-Service Program.
    
    Emmanuel Delande and Daniel Clark are supported by the Engineering and Physical Sciences Research Council (EPSRC) Platform Grant (EP/J015180/1), the MOD University Defence Research Centre on Signal Processing (UDRC) Phase 2 (EP/K014227/1).
    
    Daniel Clark wishes to thank Professor Penina Axelrad in the Aerospace Department in Boulder for supporting his Visiting Professor position through the \ac{first} programme at the University of Colorado Boulder in summer 2014.
    
    The authors would also like to thank Nicola Baresi and In-Kwan Park of the University of Colorado at Boulder and Ill\'an Amor of Universidad de Oviedo, Asturias Spain for their conversations and ideas early on for this work during the \ac{first} programme.

  \newpage
  \appendix
  \section{Proofs}  \label{sec:proof}
    \renewcommand{\IEEEQED}{\IEEEQEDopen}
    \subsection{Proof of theorem \ref{th:prediction_step}} \label{subsec:prediction_proof}
	For the sake of simplicity, the time subscripts will be omitted throughout the proof when there is no ambiguity. Also, we will denote by $\bar{p}_{\s}$ (\ac{resp} $\bar{p_{\d}}$) the function $1 - p_{\s, k}$ (\ac{resp} $1 - p_{\d, k}$). 
	\subsubsection{Predicted \ac{pgfl}}
	  Let us focus first on the \ac{pgfl} $G_{k|k-1}$ of the predicted multi-target point process $\Phi_{k|k-1}$. Each parent target in the population, represented by the prior point process $\Phi_{k-1}$, generates daughter targets in the predicted population in two ways:
	  \begin{itemize}
	    \item a daughter target stemming from the (eventual) survival of the parent target, represented by a survival point process $\Phi_{\s}$,
	    \item a population of daughter spawned from the parent target, represented by a spawning point process $\Phi_{\b}$. 
	  \end{itemize}
	  Using Eq.~\eqref{eq:superposition}, and denoting by $G_{\s}$ (\ac{resp} $G_{\b}$) the \ac{pgfl} of the survival (\ac{resp} spawning) point process, we can describe the evolution of a parent target with state $x \in \Xbo$ with a compound process with \ac{pgfl}
	  \begin{equation}
	    G_{\c}(h|x) = G_{\s}(h|x)G_{\b}(h|x), \label{eq:compound_prediction}
	  \end{equation}
	  and exploiting the Galton-Watson equation \eqref{eq:galton_watson}, we may finally write
	  \begin{subequations} \label{eq:pgfl_prediction_proof}
	    \begin{align}
	      G_{k|k-1}(h) &= G_{k-1}(G_{\c}(h|\cdot))
	      \\
	      &= G_{k-1}(G_{\s}(h|\cdot)G_{\b}(h|\cdot)).
	    \end{align}
	  \end{subequations}
	
	\subsubsection{Predicted intensity}
	  Let us now focus on the expression of the predicted intensity $\mu_{k|k-1}$. For that, let us fix an arbitrary measurable subset $B \in \Bbo_{\Xbo}$. The expression of the intensity evaluated in $B$ can be recovered from the first derivative of the \ac{pgfl} $G_{k|k-1}$ using Eq.~\eqref{eq:pgfl_derivation_setting_1}:
	  \begin{subequations}
	    \begin{align}
	      \mu_{k|k-1}(B) &= \left.\delta G_{k|k-1}(h; \ind{B})\right|_{h = 1}
	      \\
	      &= \delta \big(G_{k-1}(G_{\c}(h|\cdot)); \ind{B}\big)\big|_{h = 1}
	      \intertext{Using the definition of the \ac{pgfl} \eqref{eq:point_process_pgfl_1} then yields}
	      \mu_{k|k-1}(B) &= \delta \bigg(\int_{\Xcal} \bigg[\prod_{x \in \varphi} G_{\c}(h|x)\bigg] P_{k-1}(\d\varphi); \ind{B}\bigg) \bigg|_{h = 1}
	      \\
	      &= \int_{\Xcal} \delta \bigg(\prod_{x \in \varphi} G_{\c}(h|x); \ind{B}\bigg) \bigg|_{h = 1} P_{k-1}(\d\varphi)
	      \intertext{From the product rule \eqref{eq:product_rule} it follows that}
	      \mu_{k|k-1}(B) &= \nonumber
	      \\
	      &\!\!\!\!\int_{\Xcal} \sum_{x \in \varphi} \bigg[\delta G_{\c}(h|x; \ind{B}) \bigg|_{h = 1} \prod_{\substack{\bar{x} \in \varphi \\ \bar{x} \neq x}}\underbrace{G_{\c}(1|\bar{x})}_{=1}\bigg] P_{k-1}(\d\varphi)
	      \intertext{Using the product rule \eqref{eq:product_rule} on $G_{\c}(\cdot|x) = G_{\s}(\cdot|x)G_{\b}(\cdot|x)$ then yields}
	      \mu_{k|k-1}(B) &= \int_{\Xcal} \sum_{x \in \varphi} \bigg[\delta G_{\s}(h|x; \ind{B})\big|_{h = 1}\underbrace{G_{\b}(1|x)}_{=1} \nonumber
	      \\
	      &\hspace{30pt}+ \underbrace{G_{\s}(1|x)}_{=1}\delta G_{\b}(h|x; \ind{B})\big|_{h = 1}\bigg] P_{k-1}(\d\varphi)
	      \intertext{Using Eq.~\eqref{eq:pgfl_derivation_setting_1} we introduce the intensity $\mu_{\s}$ (\ac{resp} $\mu_{\b}$) of the survival (\ac{resp} spawning) process and we obtain:}
	      \mu_{k|k-1}(B) &= \int_{\Xcal} \sum_{x \in \varphi} \left[\mu_{\s}(B|x) + \mu_{\b}(B|x)\right] P_{k-1}(\d\varphi)
	      \intertext{Which becomes, using Campbell's theorem \cite[p. 271]{VereJones_D_2008}:}
	      \mu_{k|k-1}(B) &= \int \left[\mu_{\s}(B|x) + \mu_{\b}(B|x)\right] \mu_{k-1}(\d x).
	    \end{align}
	  \end{subequations}
	  Note that the validity of the expression of the predicted intensity above is not restricted to specific models for the prior process $\Phi_{k-1}$. As such, the construction of the predicted intensity is identical in the case of the \ac{phd} filter with spawning (see Mahler's original proof in \cite{Mahler_RPS_2003}). Let us now focus on the explicit expression of the intensity measure $\mu_{\s}$. Since the survival process is assumed Bernoulli with parameter $p_{\s}(\cdot)$ and spatial distribution $f_{\s}(\cdot|\cdot)$, we can exploit Eq.~\eqref{eq:pgfl_derivation_setting_1} to retrieve the intensity $\mu_{\s}$ through the expression of the \ac{pgfl} $G_{\s}$ given by Eq.~\eqref{eq:pgfl_bernoulli}:
	  \begin{subequations} \label{eq:bernoulli_dvpt}
	    \begin{align}
	      \mu_{\s}(B|\cdot) &= \left.\delta G_{\s}(h|\cdot; \ind{B})\right|_{h = 1}
	      \\
	      &= \delta \bigg(1 - p_{\s}(\cdot) + p_{\s}(\cdot)\int h(x)f_{\s}(\d x|\cdot); \ind{B}\bigg)\bigg|_{h = 1}
	      \\
	      &= p_{\s}(\cdot)f_{\s}(B|\cdot).
	    \end{align}
	  \end{subequations}
	  
	  Let us now focus on the explicit expression of the intensity measure $\mu_{\b}$ of the spawning process, depending on the modeling choices.
	  
	  a) Bernoulli process with parameter $p_{\b}(\cdot)$ and spatial distribution $s_{\b}(\cdot|\cdot)$:\newline    
	  Using the same construction as in Eq.~\eqref{eq:bernoulli_dvpt} we have immediately
	  \begin{equation}
	    \mu_{\b}(B|\cdot) = p_{\b}(\cdot)s_{\b}(B|\cdot).
	  \end{equation}
	  
	  b) zero-inflated Poisson process with parameter $p_{\b}(\cdot)$, rate $\lambda_{\b}(\cdot)$ and spatial distribution $s_{\b}(\cdot|\cdot)$:\newline
	  Exploiting Eq.~\eqref{eq:pgfl_bernoulli_poisson} yields
	  \begin{subequations}
	    \begin{align}
	      &\mu_{\b}(B|\cdot)\nonumber
	      \\
	      &= \left.\delta G_{\b}\left(h|\cdot; \ind{B}\right)\right|_{h = 1}
	      \\
	      &= \delta \bigg(\!\bar{p}_{\b}(\cdot)\!+\!p_{\b}(\cdot)\exp\!\bigg[\!\lambda_{\b}(\cdot)\!\bigg(\!\!\int h(x)s_{\b}(\d x|\cdot)\!-\!1 \!\bigg)\!\bigg]; \ind{B}\bigg) \bigg|_{h = 1}
	      \\
	      &= p_{\b}(\cdot)\lambda_{\b}(\cdot) \delta \bigg(\int h(x)s_{\b}(\d x|\cdot) - 1; \ind{B}\bigg)\bigg|_{h = 1} \nonumber
	      \\
	      &\hspace{80pt}\times \exp \underbrace{\bigg[\lambda_{\b}(\cdot) \bigg(\int s_{\b}(\d x|\cdot) - 1 \bigg) \bigg]}_{= 0}
	      \\
	      &= p_{\b}(\cdot)\lambda_{\b}(\cdot)s_{\b}(B|\cdot).
	    \end{align}
	  \end{subequations}
	  
	\subsubsection{Predicted cardinality}
	  Let us now focus on the expression of the predicted cardinality $\rho_{k|k-1}$. From Eq.~\eqref{eq:point_process_cardinality} the cardinality distribution of an arbitrary point process can be retrieved through its Janossy measures; let us then compute the predicted \nth-order Janossy measure $J^{(n)}_{k|k-1}$ evaluated at the neighborhood of a  collection of $n$ arbitrary points $y_1,\ldots,y_n$. Using Eq.~\eqref{eq:pgfl_derivation_setting_0} yields
	  \begin{subequations}
	    \begin{align}
	      &J_{k|k-1}^{(n)}(\d(y_1,\ldots,y_n)) \nonumber
	      \\
	      &= \left.\delta^n G_{k|k-1}(h;\ind{\d y_1},\ldots,\ind{\d y_n}) \right|_{h=0}
	      \\
	      &= \left.\delta^n (G_{k-1}(G_{\c}(h|\cdot)); \ind{\d y_1},\ldots,\ind{\d y_n})\right|_{h=0}
	      \intertext{Applying the general chain rule \eqref{eq:chain_rule} then gives}	
	      &J_{k|k-1}^{(n)}(\d(y_1,\ldots,y_n)) \nonumber
	      \\
	      &= \!\!\sum_{\pi \in \Pi_n} \!\!\delta^{|\pi|} G_{k-1} \left(\!\!G_{\c}(h|\cdot); \left(\delta^{|\omega|} G_{\c}(h|\cdot; (\ind{\d y_i} )_{i \in \omega})\right)_{\omega \in \pi} \right)\!\!\bigg|_{h=0}.
	    \end{align}
	  \end{subequations}
	  Developing the predicted \ac{pgfl} $G_{k-1}$ through Janossy measures with Eq.~\eqref{eq:point_process_description} then gives
	  \begin{align}
	    &J_{k|k-1}^{(n)}(\d(y_1,\ldots,y_n)) =\nonumber
	    \\
	    &\sum_{\pi \in \Pi_n} \sum_{m\geq |\pi|} \frac{1}{(m-|\pi|)!} \int_{\Xbo^m} \prod_{i=1}^{|\pi|} \delta^{|\omega_i|} G_{\c}(h|x_i; (\ind{\d y_j})_{j \in \omega_i}) \bigg|_{h=0} \nonumber
	    \\
	    &\hspace{80pt}\times \prod_{i=|\pi|+1}^m G_{\c}(0|x_i) J_{k-1}^{(m)}(\d(x_1, \ldots, x_m)).
	  \end{align}
	  Since the prior process is assumed \ac{iid}, we can substitute the expression given by Eq.~\eqref{eq:janossy_iid} to the prior Janossy densities $J_{k-1}^{(m)}$ and obtain
	  \begin{multline}
	    J_{k|k-1}^{(n)}(\d(y_1,\ldots,y_n))	      
	    \\
	    = \sum_{\pi \in \Pi_n} \sum_{m\geq |\pi|} \frac{m!}{(m-|\pi|)!} \rho(m)C_{\pi}(\d(y_1,\ldots,y_n)), \label{eq:janossy_generic_red}
	  \end{multline}
	  where 
	  \begin{subequations} \label{eq:janossy_generic}
	    \begin{align}
	      &C_{\pi}(\d(y_1,\ldots,y_n))\nonumber
	      \\
	      &= \idotsint \prod_{i=1}^{|\pi|} \left.\delta^{|\omega_i|} G_{\c}(h|x_i; (\ind{\d y_j})_{j \in \omega_i})\right|_{h=0} \nonumber
	      \\
	      &\hspace{80pt}\times \prod_{i=|\pi|+1}^m G_{\c}(0|x_i) \prod_{i = 1}^m s(\d x_i)
	      \\
	      &= \bigg(\int G_{\c}(0|x) s(\d x)\bigg)^{m-|\pi|}\nonumber
	      \\
	      &\hspace{12pt}\times  \prod_{\omega \in \pi} \bigg(\int \delta^{|\omega|} \left.G_{\c}(h|x; (\ind{\d y_i})_{i \in \omega}) \right|_{h=0} s(\d x)\bigg) \label{eq:janossy_generic_end}
	    \end{align}
	  \end{subequations}
	  Recall from Eq.~\eqref{eq:compound_prediction} that $G_{\c}(h|x) = G_{\s}(h|x)G_{\b}(h|x)$; using the product rule \eqref{eq:product_rule} on Eq.~\eqref{eq:janossy_generic_end} then yields
	  \begin{align} \label{eq:janossy_generic_2}
	    &C_{\pi}(\d(y_1,\ldots,y_n)) = \bigg(\int G_{\s}(0|x) G_{\b}(0|x)s(\d x)\bigg)^{m-|\pi|} \nonumber
	    \\
	    &\times \prod_{\omega \in \pi} \bigg(\int \sum_{\nu \subseteq \omega} \delta^{|\nu|} G_{\s}(h|x; (\ind{\d y_i})_{i \in \nu}) \bigg|_{h=0} \nonumber
	    \\
	    &\hspace{35pt}\delta^{|\omega| - |\nu|} G_{\b}(h|x; (\ind{\d y_i})_{i \in \omega \setminus \nu}) \bigg|_{h=0} s(\d x)\bigg). 
	  \end{align}
	  Now, from the derivation shown in Eq.~\eqref{eq:bernoulli_dvpt}, we see that:
	  \begin{equation} \label{eq:bernoulli_dvpt_2}
	    \delta^{|\nu|} G_{\s}(h|x; (\ind{\d y_i})_{i \in \nu}) \big|_{h=0} =
	    \begin{cases}
	      1 - p_{\s}(x), &\nu = \emptyset,
	      \\
	      p_{\s}(x)f_{\s}(\d y_i|x), &\nu = \{i\},
	      \\
	      0, &|\nu| > 1.
	    \end{cases}
	  \end{equation}
	  Therefore, Eq.~\eqref{eq:janossy_generic_2} simplifies as follows:
	  \begin{align}
	    &C_{\pi}(\d(y_1,\ldots,y_n)) = \bigg(\int \bar{p}_{\s}(x)G_{\b}(0|x)s(\d x)\bigg)^{m-|\pi|} \nonumber
	    \\
	    &\times \prod_{\omega \in \pi} \bigg(\int \bar{p}_{\s}(x) \delta^{|\omega|} G_{\b}(h|x; (\ind{\d y_i})_{i \in \omega}) \big|_{h=0}s(\d x) \nonumber
	    \\
	    & + \int \sum_{\ind{\d y_i} \in \omega} p_{\s}(x)f_{\s}(\d y_i|x) \nonumber
	    \\
	    &\times  \delta^{|\omega| - 1}G_{\b}(h|x; (\ind{\d y_j})_{j \in \omega \setminus \{i\}}) \big|_{h=0} s(\d x)\bigg). \label{eq:janossy_generic_3}
	  \end{align}
	  We shall now detail the expression of Eq.~\eqref{eq:janossy_generic_3} depending on the modeling choices for the spawning process.
	  
	  a) Bernoulli process with parameter $p_{\s}(\cdot)$ and spatial distribution $f_{\s}(\cdot|\cdot)$:\newline    
	  We may draw similar results from the derivation shown in Eq.~\eqref{eq:bernoulli_dvpt_2}:
	  \begin{equation}
	    \delta^{|\nu|} G_{\b}(h|x; (\ind{\d y_j})_{j \in \nu})\big|_{h=0} =
	    \begin{cases}
	      1 - p_{\b}(x), &\nu = \emptyset,
	      \\
	      p_{\b}(x)s_{\b}(\d y_j|x), &\nu = \{j\},
	      \\
	      0, &|\nu| > 1.
	    \end{cases}
	  \end{equation}
	  Therefore, Eq.~\eqref{eq:janossy_generic_3} simplifies as follows
	  \begin{align}
	    &C_{\pi}(\d(y_1,\ldots,y_n)) = \bigg(\int \bar{p}_{\s}(x)\bar{p}_{\b}(x)s(\d x)\bigg)^{m-|\pi|} \nonumber
	    \\
	    &\times \prod_{\{i\} \in \pi} \bigg(\int \bar{p}_{\s}(x)p_{\b}(x)s_{\b}(\d y_i|x)s(\d x) \nonumber
	    \\
	    &\hspace{93pt}+ \int \bar{p}_{\b}(x)p_{\s}(x)f_{\s}(\d y_i|x)s(\d x)\bigg) \nonumber
	    \\
	    &\times \prod_{\{i, j\} \in \pi} \bigg(\int p_{\s}(x)p_{\b}(x)f_{\s}(\d y_i|x)s_{\b}(\d y_j|x)s(\d x) \nonumber
	    \\
	    &\hspace{55pt}+ \int p_{\s}(x)p_{\b}(x)f_{\s}(\d y_j|x)s_{\b}(\d y_i|x)s(\d x)\bigg) \nonumber
	    \\
	    &\times \prod_{\substack{\omega \in \pi \\ |\omega| > 2}} 0. \label{eq:janossy_bernoulli}
	  \end{align}
	  Substituting Eq.~\eqref{eq:janossy_bernoulli} to Eq.~\eqref{eq:janossy_generic_red}, we may finally retrieve the scalar $\rho_{k|k-1}(n)$ through Eq.~\eqref{eq:point_process_cardinality}:
	  \begin{subequations}
	    \begin{align} 
	      &\rho_{k|k-1}(n) \nonumber
	      \\
	      &= \frac{1}{n!} \int_{\Xbo^n} J_{k|k-1}^{(n)}(\d(y_1,\ldots,y_n))
	      \\
	      &= \sum_{\pi \in \Pi_n} \sum_{m\geq |\pi|} \frac{m!}{n!(m-|\pi|)!} \rho(m) b_0^{m-|\pi|} \prod_{\omega \in \pi} b_{|\omega|}, \label{eq:BernCardRho}
	    \end{align}
	  \end{subequations}
	  where the coefficients $b_i$ are defined by
	  \begin{equation}
	    b_i =
	    \begin{cases}
	      \int \bar{p}_{\s}(x)\bar{p}_{\b}(x)s(\d x), &i = 0,
	      \\
	      \int \left[p_{\s}(x)\bar{p}_{\b}(x) + \bar{p}_{\s}(x)p_{\b}(x)\right]s(\d x), &i = 1,
	      \\
	      2\int p_{\s}(x)p_{\b}(x) s(\d x), &i = 2,
	      \\
	      0, &i > 2.
	    \end{cases}
	  \end{equation}
	  Using the definition of the Bell polynomial \eqref{eq:BellPoly} then yields the desired result.
	  
	  b) zero-inflated Poisson process with parameter $p_{\b}(\cdot)$, rate $\lambda_{\b}(\cdot)$, and spatial distribution $s_{\b}(\cdot|\cdot)$:\newline
	  Applying the chain rule \eqref{eq:chain_rule} to the \ac{pgfl} \eqref{eq:pgfl_bernoulli_poisson} yields
	  \begin{multline}
	    \delta^{|\nu|} G_{\b}(h|x; (\ind{\d y_j})_{j \in \nu})\big|_{h=0} =
	    \\
	    \begin{cases}
	      1 - p_{\b}(x) + p_{\b}(x)e^{-\lambda_{\b}(x)}, &\nu = \emptyset,
	      \\
	      p_{\b}(x)e^{-\lambda_{\b}(x)}\lambda_{\b}(x)^{|\nu|}\prod_{j \in \nu} s_{\b}(\d y_j|x), &|\nu| > 0.
	    \end{cases}
	  \end{multline}
	  Therefore, Eq.~\eqref{eq:janossy_generic_3} simplifies as follows
	  \begin{align}
	    &C_{\pi}(\d(y_1,\ldots,y_n)) =\nonumber
	    \\
	    &\bigg(\int \bar{p}_{\s}(x)\bigg[\bar{p}_{\b}(x) + p_{\b}(x)e^{-\lambda_{\b}(x)}\bigg]s(\d x)\bigg)^{m-|\pi|} \nonumber
	    \\
	    &\times \prod_{\{i\} \in \pi} \bigg(\int \bar{p}_{\s}(x)p_{\b}(x)e^{-\lambda_{\b}(x)}\lambda_{\b}(x)s_{\b}(\d y_i|x)s(\d x)  \nonumber
	    \\
	    &\hspace{30pt}+ \int p_{\s}(x)(\bar{p}_{\b}(x) + p_{\b}(x)e^{-\lambda_{\b}(x)}) f_{\s}(\d y_i|x)s(\d x)\bigg) \nonumber
	    \\
	    &\times \prod_{\substack{\omega \in \pi \\ |\omega| > 1}} \bigg(\int \bar{p}_{\s}(x)p_{\b}(x)e^{-\lambda_{\b}(x)}\lambda_{\b}(x)^{|\omega|}\bigg[\prod_{i \in \omega} s_{\b}(\d y_i|x)\bigg] s(\d x) \nonumber
	    \\
	    &\hspace{30pt}+ \int p_{\s}(x)p_{\b}(x)e^{-\lambda_{\b}(x)}\lambda_{\b}(x)^{|\omega|-1} \nonumber
	    \\
	    &\hspace{55pt}\times \sum_{i \in \omega} f_{\s}(\d y_i|x) \bigg[\prod_{\substack{j \in \omega \\ j \neq i}} s_{\b}(\d y_j|x)\bigg] s(\d x)\bigg). \label{eq:janossy_bernoulli_poisson}
	  \end{align}
	  Substituting Eq.~\eqref{eq:janossy_bernoulli_poisson} to Eq.~\eqref{eq:janossy_generic_red}, we may finally retrieve the scalar $\rho_{k|k-1}(n)$ through Eq.~\eqref{eq:point_process_cardinality}:
	  \begin{subequations}
	    \begin{align} 
	      &\rho_{k|k-1}(n) \nonumber
	      \\
	      &= \frac{1}{n!} \int_{\Xbo^n} J_{k|k-1}^{(n)} (\d(y_1,\ldots,y_n))
	      \\
	      &= \sum_{\pi \in \Pi_n} \sum_{m\geq |\pi|} \frac{m!}{n!(m-|\pi|)!} \rho(m) b_0^{m-|\pi|} \prod_{\omega \in \pi} b_{|\omega|} \label{eq:ZifCardRho}
	    \end{align}
	  \end{subequations}
	  where the coefficients $b_i$ are defined by
	  \begin{equation}
	    b_i =
	    \begin{cases}
	      \int \bar{p}_{\s}(x)\left[\bar{p}_{\b}(x) + p_{\b}(x)e^{-\lambda_{\b}(x)}\right]s(\d x),
	      \\
	      \hspace{173pt}i = 0,
	      \\
	      \int \big[\bar{p}_{\s}(x)p_{\b}(x)e^{-\lambda_{\b}(x)}\lambda_{\b}(x)
	      \\
	      \hspace{20pt}+ p_{\s}(x)\left[\bar{p}_{\b}(x) + p_{\b}(x)e^{-\lambda_{\b}(x)}\right]\big]s(\d x),
	      \\
	      \hspace{173pt}i = 1,
	      \\
	      \int p_{\b}(x)\lambda_{\b}^{i-1}(x) e^{-\lambda_{\b}(x)}
	      \\
	      \hspace{64pt}\left[\bar{p}_{\s}(x)\lambda_{\b}(x) + i p_{\s}(x)\right]s(\d x),
	      \\
	      \hspace{173pt}i \geq 2.
	    \end{cases}
	  \end{equation}
	  Using the definition of the Bell polynomial \eqref{eq:BellPoly} then yields the desired result.   
  
  \ifCLASSOPTIONcaptionsoff
    \newpage
  \fi

  \bibliographystyle{IEEEtran}
  \bibliography{references}   

\begin{thebibliography}{10}
\providecommand{\url}[1]{#1}
\csname url@samestyle\endcsname
\providecommand{\newblock}{\relax}
\providecommand{\bibinfo}[2]{#2}
\providecommand{\BIBentrySTDinterwordspacing}{\spaceskip=0pt\relax}
\providecommand{\BIBentryALTinterwordstretchfactor}{4}
\providecommand{\BIBentryALTinterwordspacing}{\spaceskip=\fontdimen2\font plus
\BIBentryALTinterwordstretchfactor\fontdimen3\font minus
  \fontdimen4\font\relax}
\providecommand{\BIBforeignlanguage}[2]{{%
\expandafter\ifx\csname l@#1\endcsname\relax
\typeout{** WARNING: IEEEtran.bst: No hyphenation pattern has been}%
\typeout{** loaded for the language `#1'. Using the pattern for}%
\typeout{** the default language instead.}%
\else
\language=\csname l@#1\endcsname
\fi
#2}}
\providecommand{\BIBdecl}{\relax}
\BIBdecl

\bibitem{Fortmann_TE_1983}
T.~E. Fortmann, Y.~Bar-Shalom, and M.~Scheffe, ``{Sonar tracking of multiple
  targets using joint probabilistic data association},'' \emph{Oceanic
  Engineering, IEEE Journal of}, vol.~8, no.~3, pp. 173--184, Jul 1983.

\bibitem{Reid_D_1979}
D.~Reid, ``{An Algorithm for Tracking Multiple Targets},'' \emph{Automatic
  Control, IEEE Transactions on}, vol.~24, no.~6, pp. 843--854, Dec. 1979.

\bibitem{Mahler_RPS_2007_3}
R.~P.~S. Mahler, \emph{{Statistical Multisource-Multitarget Information
  Fusion}}.\hskip 1em plus 0.5em minus 0.4em\relax Artech House, 2007.

\bibitem{Mahler_RPS_2014}
------, \emph{{Advances in Statistical Multisource-Multitarget Information
  Fusion}}.\hskip 1em plus 0.5em minus 0.4em\relax Artech House, 2014.

\bibitem{Mahler_RPS_2003}
------, ``{Multitarget Bayes Filtering via First-Order Multitarget Moments},''
  \emph{Aerospace and Electronic Systems, IEEE Transactions on}, vol.~39,
  no.~4, pp. 1152--1178, Oct. 2003.

\bibitem{Mahler_RPS_2007}
------, ``{PHD Filters of Higher Order in Target Number},'' \emph{Aerospace and
  Electronic Systems, IEEE Transactions on}, vol.~43, no.~4, pp. 1523--1543,
  Oct. 2007.

\bibitem{Swartwout_M_2011}
M.~Swartwout, ``{A brief history of rideshares (and attack of the CubeSats)},''
  in \emph{{Aerospace Conference, 2011 IEEE}}, mar 2011, pp. 1--15.

\bibitem{Swartwout_M_2012}
------, ``{A statistical survey of rideshares (and attack of the CubeSats, part
  deux)},'' in \emph{{Aerospace Conference, 2012 IEEE}}, mar 2012, pp. 1--7.

\bibitem{Anselmo_L_2009}
L.~Anselmo and C.~Pardini, ``{Analysis of the consequences in low Earth orbit
  of the collision between Cosmos 2251 and Iridium 33},'' in \emph{{21st
  International Symposiumon Space Flight Dynamics}}, vol. 294, 2009.

\bibitem{Johnson_NL_2008}
N.~L. Johnson, E.~Stansbery, J.-C. Liou, M.~Horstman, C.~Stokely, and
  D.~Whitlock, ``{The characteristics and consequences of the break-up of the
  Fengyun-1C spacecraft},'' \emph{Acta Astronautica}, vol.~63, no.~1, pp.
  128--135, 2008.

\bibitem{Jones_BA_2015_1}
B.~A. Jones, D.~S. Bryant, B.-T. Vo, and B.-N. Vo, ``{Challenges of
  Multi-Target Tracking for Space Situational Awareness},'' ser. {Information
  Fusion, Proceedings of the 18th International Conference on}, July 2015.

\bibitem{Jones_BA_2014}
B.~A. Jones, S.~Gehly, and P.~Axelrad, ``{Measurement-based Birth Model for a
  Space Object Cardinalized Probability Hypothesis Density Filter},''
  \emph{AIAA/AAS Astrodynamics Specialist Conference}, Aug. 2014.

\bibitem{Lundgren_M_2013_1}
M.~Lundgren, L.~Svensson, and L.~Hammarstrand, ``{A CPHD filter for tracking
  with spawning models},'' \emph{Selected Topics in Signal Processing, IEEE
  Journal of}, vol.~7, no.~3, pp. 496--507, 2013.

\bibitem{Stoyan_D_1995}
D.~Stoyan, W.~S. Kendall, and J.~Mecke, \emph{{Stochastic geometry and its
  applications}}, 2nd~ed.\hskip 1em plus 0.5em minus 0.4em\relax Wiley, Sep.
  1995.

\bibitem{Vo_BN_2005}
B.-N. Vo, S.~Singh, and A.~Doucet, ``{Sequential Monte Carlo methods for
  Multi-target Filtering with Random Finite Sets},'' \emph{Aerospace and
  Electronic Systems, IEEE Transactions on}, vol.~41, no.~4, pp. 1224--1245,
  Oct. 2005.

\bibitem{VereJones_D_2003}
D.~Vere-Jones and D.~J. Daley, \emph{{An Introduction to the Theory of Point
  Processes}}, 2nd~ed., ser. {Statistical Theory and Methods}, D.~Vere-Jones
  and D.~J. Daley, Eds.\hskip 1em plus 0.5em minus 0.4em\relax Springer Series
  in Statistics, 2003, vol.~1.

\bibitem{Delande_E_2014_4}
E.~Delande, M.~Uney, J.~Houssineau, and D.~E. Clark, ``{Regional Variance for
  Multi-Object Filtering},'' \emph{IEEE Transactions on Signal Processing},
  vol.~62, no.~13, pp. 3415 -- 3428, Jul. 2014.

\bibitem{Srinivasan_SK_2003}
S.~K. Srinivasan and A.~Vijayakumar, \emph{{Point Processes and Product
  Densities}}.\hskip 1em plus 0.5em minus 0.4em\relax Alpha Science Int'l Ltd.,
  2003.

\bibitem{Moyal_JE_1962}
J.~E. Moyal, ``{The General Theory of Stochastic Population Processes},''
  \emph{Acta Mathematica}, vol. 108, no.~1, pp. 1--31, Dec. 1962.

\bibitem{Watson_HW_1875}
H.~W. Watson and F.~Galton, ``{On the Probability of the Extinction of
  Families},'' \emph{Journal of the Anthropological Institute of Great
  Britain}, vol.~4, pp. 138-----144, 1875.

\bibitem{Bernhard_P_2006}
P.~Bernhard, ``{Chain differentials with an application to the mathematical
  fear operator},'' \emph{Nonlinear Analysis}, vol.~62, pp. 1225--1233, 2005.

\bibitem{Clark_DE_2013_2}
D.~E. Clark and J.~Houssineau, ``{Fa{\`a} di Bruno's formula and spatial
  cluster modelling},'' \emph{Spatial Statistics}, vol.~6, pp. 109--117, 2013.

\bibitem{Clark_DE_2013_3}
------, ``{Fa{\`a} di Bruno's formula for chain differentials},''
  \emph{arXiv:1310.2833}, 2013.

\bibitem{FaaDiBruno_F_1855}
F.~{Fa{\`a} di Bruno}, ``{Sullo Sviluppo delle Funzioni},'' \emph{Annali di
  Scienze Matematiche e Fisiche}, vol.~6, pp. 479--480, 1855.

\bibitem{Clark_DE_2015_1}
D.~E. Clark, J.~Houssineau, and E.~Delande, ``{A few calculus rules for chain
  differentials},'' \emph{arXiv:1506.08626}, 2015.

\bibitem{Srinivasan_SK_1973}
S.~K. Srinivasan, \emph{{Stochastic Point Processes and Their
  Applications}}.\hskip 1em plus 0.5em minus 0.4em\relax Griffin's Statistical
  Monographs and Courses, 1973.

\bibitem{Vo_BT_2007}
B.-T. Vo, B.-N. Vo, and A.~Cantoni, ``{Analytic Implementations of the
  Cardinalized Probability Hypothesis Density Filter},'' \emph{Signal
  Processing, IEEE Transactions on}, vol.~55, pp. 3553--3567, 2007.

\bibitem{Lambert_D_1992}
D.~Lambert, ``{Zero-Inflated Poisson Regression, with an Application to Defects
  in Manufacturing},'' \emph{Technometrics}, vol.~34, no.~1, pp. 1--14, feb
  1992.

\bibitem{Charalambides_C_2002}
C.~A. Charalambides, \emph{{Enumerative combinatorics}}.\hskip 1em plus 0.5em
  minus 0.4em\relax CRC Press, 2002.

\bibitem{Cvijovic_D_2011_1}
D.~Cvijovi{\'c}, ``{New identities for the partial Bell polynomials},''
  \emph{Applied Mathematics Letters}, vol.~24, no.~9, pp. 1544--1547, 2011.

\bibitem{Vo_BN_2006_2}
B.-N. Vo and W.-K. Ma, ``{The Gaussian Mixture Probability Hypothesis Density
  Filter},'' \emph{Signal Processing, IEEE Transactions on}, vol.~54, no.~11,
  pp. 4091--4104, Nov. 2006.

\bibitem{Schuhmacher_D_2008}
D.~Schuhmacher, B.-N. Vo, and B.-T. Vo, ``{A Consistent Metric for Performance
  Evaluation of Multi-object filters},'' \emph{Signal Processing, IEEE
  Transactions on}, vol.~56, no.~8, pp. 3447--3457, Aug. 2008.

\bibitem{Gibbs_AL_2002_1}
A.~L. Gibbs and F.~E. Su, ``{On Choosing and Bounding Probability Metrics},''
  \emph{International Statistical Review}, vol.~70, no.~3, pp. 419--435, 2002.

\bibitem{VereJones_D_2008}
D.~Vere-Jones and D.~J. Daley, \emph{{An Introduction to the Theory of Point
  Processes}}, 2nd~ed., ser. {Statistical Theory and Methods}, D.~Vere-Jones
  and D.~J. Daley, Eds.\hskip 1em plus 0.5em minus 0.4em\relax Springer Series
  in Statistics, 2008, vol.~2.

\end{thebibliography}

%
%
%
%
%

\end{document}